\def\dir{./}
\documentclass[useAMS,usenatbib]{\dir mn2e}
\usepackage[fleqn]{amsmath}
\usepackage{amssymb}
\usepackage[super]{nth}
\usepackage{natbib}
\usepackage{graphicx}
\usepackage{float}
\usepackage{mathrsfs}
\usepackage{mathtools}
\usepackage{multirow}
\usepackage[usenames,dvipsnames]{color}
\usepackage{\dir aas_macros}
\usepackage{comment}
\usepackage{subfigure}
\usepackage{hyperref}
\usepackage{mathrsfs}
\usepackage{epstopdf, dcolumn}
\DeclareGraphicsExtensions{.pdf}
\usepackage{wasysym}
\usepackage{dashrule}
\usepackage{xcolor}
\usepackage{arydshln}
\usepackage{threeparttable}
\usepackage[export]{adjustbox}
\usepackage[figuresleft]{rotating}

\newlength\replength
\newcommand\repfrac{.33}

\newcommand\rulewidth{.6pt}
\newcommand\tdashfill[1][\repfrac]{\cleaders\hbox to \replength{%
		\smash{\rule[\arraystretch\ht\strutbox]{\repfrac\replength}{\rulewidth}}}\hfill}

\newcommand\tdotfill[1][\repfrac]{\cleaders\hbox to \replength{%
		\smash{\raisebox{\arraystretch\dimexpr\ht\strutbox-.1ex\relax}{.}}}\hfill}

\newcommand{\appropto}{\mathrel{\vcentre{
			\offinterlineskip\halign{\hfil$##$\cr
				\propto\cr\noalign{\kern2pt}\sim\cr\noalign{\kern-2pt}}}}}

\newcommand{\cmfast}{\textsc{\small 21cmFAST}}

\newcommand\lsim{\mathrel{\rlap{\lower4pt\hbox{\hskip1pt$\sim$}}
        \raise1pt\hbox{$<$}}}
\newcommand\gsim{\mathrel{\rlap{\lower4pt\hbox{\hskip1pt$\sim$}}
        \raise1pt\hbox{$>$}}}
\newcommand{\Rom}[1]{\uppercase\expandafter{\romannumeral #1}}
\newcommand{\rom}[1]{\lowercase\expandafter{\romannumeral #1}}
\newcommand{\htwo}{\mathrm{H}_2}
\newcommand{\hone}{\mathrm{H}\textsc{i}}
\newcommand{\hii}{\mathrm{H}\textsc{ii}}
\newcommand{\lya}{{Lyman-$\alpha$}}
\newcommand{\mvir}{M_{\rm vir}}
\newcommand{\msol}{{\rm M}_\odot}
\newcommand{\rmd}{{\rm d}}
\newcommand{\omegam}{\Omega_{\rm m}}

\defcitealias{Park2019MNRAS.484..933P}{P19}

\begin{document}

\title[minihaloes in 21cmFAST]{A tale of two sites - \Rom{1}: Inferring the properties of minihalo-hosted galaxies from current observations}

\author[Qin et al.]{Yuxiang Qin$^{1}$\thanks{E-mail: Yuxiang.L.Qin@gmail.com}, Andrei Mesinger$^1$, Jaehong Park$^1$, Bradley Greig$^{2,3}$, and \newauthor{Julian B.~Mu\~noz$^4$}\\
	$^{1}$Scuola Normale Superiore, Piazza dei Cavalieri 7, I-56126 Pisa, Italy\\
	$^{2}$School of Physics, University of Melbourne, Parkville, VIC 3010, Australia\\
	$^{3}$ARC Centre of Excellence for All Sky Astrophysics in 3 Dimensions (ASTRO 3D)\\
	$^{4}$Department of Physics, Harvard University, 17 Oxford St., Cambridge, MA 02138, USA
}
\maketitle
\label{firstpage}

\begin{abstract}
The very first galaxies that started the cosmic dawn likely resided in so-called ``minihaloes’’, with masses of ${\sim}10^5$--$10^8{\msol}$, accreting their gas from the intergalactic medium through H$_2$ cooling.  Such molecularly cooled galaxies (MCGs) mostly formed in pristine environments, hosted massive, metal-free stars, and were eventually sterilized by the build-up of a disassociating (Lyman--Werner; LW) background.  Therefore, their properties might be very different from the galaxies we see in the later Universe.  Although MCGs are probably too faint to be observed directly, we could nevertheless infer their properties from the imprint they leave in the cosmic 21-cm signal. Here we quantify this imprint by extending the public simulation code {\cmfast} to allow for a distinct population of MCGs.  We allow MCGs to have different properties from other galaxies, including unique scaling relations for their stellar-to-halo mass ratios, ionizing escape fractions, and spectral energy distributions.  We track inhomogeneous recombinations, disassociative LW feedback, and photoheating from reionization. After demonstrating how MCGs can shape the 21-cm signal, we explore to what extent current observations can already place constraints on their properties.  The cosmic microwave background optical depth from {\it Planck} sets an upper limit on the product of the ionizing escape fraction and the stellar mass in MCGs.  When including also the timing of the putative EDGES absorption signal, we find an additional strong degeneracy between the stellar mass and the X-ray luminosity of MCGs.  If proven to be of cosmic origin, the timing of the EDGES signal would have been set by MCGs.
\end{abstract}


\section{Introduction}
The hyperfine spin-flip transition of neutral hydrogen, corresponding to a photon with a wavelength of 21 cm, promises to revolutionize our understanding of the first billion years of the Universe.
The cosmic 21-cm signal is typically expressed as the brightness temperature contrast of the cosmic hydrogen against the cosmic microwave background (CMB), at a redshifted frequency $\nu$ and spatial position ${\bf r}$ (e.g. \citealt{Furlanetto2006PhR...433..181F})
\begin{equation}\label{eq:dtb_define}
	\begin{split}
		&\delta T_\mathrm{b}\left(\nu, {\bf r}\right) = \left(T_\mathrm{S} - T_\mathrm{CMB}\right)
		\left({1{-}e^{{-}\tau_\mathrm{\nu_0}}}\right){\left(1+z\right)}^{-1} \\
		&\approx 20\mathrm{mK} \left(1{-}\frac{\mathrm{T_\mathrm{CMB}}}{T_\mathrm{S}}\right) 
		\frac{x_{\mathrm{HI}} \left(1+\delta\right)}{1+\frac{\mathrm{d}v_r}{\mathrm{d}r}/H} 
		\sqrt{\frac{1{+}z}{10}\frac{0.15}{\Omega_\mathrm{m}h^2}}\frac{\Omega_\mathrm{b}h^2}{0.023},
	\end{split}
\end{equation}
where $T_\mathrm{CMB}$ is the CMB temperature; $T_{\rm S}$ is the spin temperature denoting the relative level populations of the hyperfine transition; {\color{black}$H$ is the Hubble constant at redshift $z$;} and $\tau_\mathrm{\nu_0}$ is the optical depth and is a function of the spin temperature, neutral hydrogen fraction ($x_{\hone}$), local overdensity ($\delta \equiv \rho_{\rm b}/\bar{\rho}_{\rm b} - 1$ with $\rho_{\rm b}$ and $\bar{\rho}_{\rm b}$ being the baryonic density and its cosmic mean, respectively), and the line-of-sight velocity gradient (d$v_r$/d$r$), as well as cosmological parameters, such as the present baryon and matter abundances, $\Omega_{\mathrm{b}}$ and $\Omega_{\mathrm{m}}$, and the Hubble constant, $h$.

In addition to physical cosmology, the signal is sensitive to the ionization and thermal state of the intergalactic medium (IGM), which are governed by the ionizing, X-ray and soft UV radiation fields during the cosmic dawn (CD) and subsequent epoch of reionization (EoR).  These radiation fields are established by stars and black holes inside
the first generations of galaxies (though exotic sources such as dark matter annihilation or primordial black holes might contribute; e.g. \citealt{Evoli2014JCAP...11..024E,Lopez-Honorez2016JCAP...08..004L,Hektor2018PhRvD..98b3503H}).
Thus, the cosmic 21-cm signal encodes the properties of unseen galaxies during the first billion years.
Current interferometers, including the Low-Frequency Array (LOFAR\footnote{\url{http://www.lofar.org/}}; \citealt{vanHaarlem2013A&A...556A...2V,Patil2017ApJ...838...65P}) and the Murchison Widefield Array (MWA\footnote{\url{http://www.mwatelescope.org/}}; \citealt{Tingay2013PASA...30....7T,Beardsley2016ApJ...833..102B}), are aiming for a statistical detection of the EoR; however, next-generation instruments, such as the Hydrogen Epoch of Reionization Arrays (HERA\footnote{\url{http://reionization.org/}}; \citealt{DeBoer2017PASP..129d5001D,Kohn2019ApJ...882...58K}) and the Square Kilometre Array (SKA\footnote{\url{https://www.skatelescope.org/}}; \citealt{Mellema2013ExA....36..235M,Koopmans2015aska.confE...1K}), are expected to characterize the topology of the CD, allowing us to indirectly study the very first galaxies out to $z\lsim20$--30.

The first generations of galaxies are expected to reside in so-called minihaloes, with virial temperatures below $\sim10^4$K.  At these low temperatures, cooling through atomic hydrogen (${\hone}$) and helium (He) is inefficient; therefore, minihaloes obtain gas from the IGM and  {\color{black}cool} predominately through molecular hydrogen (${\htwo}$) cooling (e.g. \citealt{Haiman1996ApJ...467..522H,Haiman1997ApJ...476..458H,Yoshida2003ApJ...598...73Y,Yoshida2006}). As they form out of pristine (unpolluted) gas, these minihaloes or molecular-cooling galaxies (MCGs) are expected to host metal-free (so-called PopIII) stars and associated remnants \citep{Wise2012ApJ...745...50W,Xu2016ApJ...833...84X}. Their shallow potential wells suggest that they are sensitive to supernova and photoheating feedback \citep{Haiman2000ApJ...534...11H,Wise2007,Kimm2016}.  Moreover, star formation inside MCGs is expected to be transient since an H$_2$-disassociating (Lyman--Werner; LW) background becomes established soon after the first stars form (e.g. \citealt{Johnson2007ApJ...665...85J,Ahn2009ApJ...695.1430A,Holzbauer2012MNRAS.419..718H,Fialkov2013MNRAS.432.2909F,Jaacks2018,Schauer2019MNRAS.484.3510S}).  Because of this uniqueness (pristine environment, top-heavy initial mass function (IMF), transient star formation, peculiar energetics), it is doubtful that the typical properties of MCGs can be adequately captured by simply extending the scaling relations inferred from observations of their massive counterparts at lower redshifts (e.g. \citealt{Mirocha2019MNRAS.483.1980M, Mebane2019arXiv191010171M}).

Unfortunately, studying MCGs through direct observations is unlikely in the foreseeable future.  Most are expected to have UV magnitudes in the range of $M_{\rm UV}{\sim}$-5 to -12 (e.g. \citealt{OShea2015ApJ...807L..12O,Xu2016ApJ...833...84X}), below the observational limit of even the next-generation infrared instrument, the {\it James Webb Space Telescope} ({\it JWST}; \citealt{Gardner2006SSRv..123..485G}).  Gravitational lensing has allowed us to push UV luminosity functions (LFs) down to $M_{\rm UV}{\sim}{-}12.5$ at $z\approx6$ (e.g. \citealt{Bouwens2017ApJ...843..129B,Livermore2017ApJ...835..113L,Atek2018MNRAS.479.5184A,Ishigaki2018ApJ...854...73I}).  However, even if the associated large systematic uncertainties can be mitigated, MCGs seem unlikely to persist at these low redshifts and few if any might be found in the effective lensing volume (e.g. \citealt{Atek2018MNRAS.479.5184A}).

Thankfully, these transient first galaxies leave an imprint in the timing and topology of the 21-cm signal.  For example, massive stars in MCGs could be responsible for a tail in the reionization history extending towards high redshifts (e.g. \citealt{Ahn2012ApJ...756L..16A,Visbal2015MNRAS.453.4456V,Miranda2017MNRAS.467.4050M}), and imprint more small-scale structure in the reionization topology (e.g. \citealt{Mesinger2012MNRAS.422.1403M, Koh2018MNRAS.474.3817K}), while the neutral gas inside minihaloes can act as ionizing photon sinks, delaying reionization and further affecting the topological features (e.g. \citealt{Ciardi2006MNRAS.366..689C,McQuinn2007MNRAS.377.1043M}).
MCGs could play an even more prominent role in the timing and morphology of the earlier epochs when X-rays and soft UV photons drive the 21-cm signal (e.g. \citealt{Fialkov2013MNRAS.432.2909F, Fialkov2014MNRAS.445..213F, Mirocha2017,Munoz2019PhRvD.100f3538M,Mebane2019arXiv191010171M}).  Indeed, these early epochs recently received attention because of the claimed detection of the globally averaged 21-cm absorption feature at $z {\sim}17$ by the Experiment to Detect the Global Epoch of Reionization Signature (EDGES; \citealt{Bowman2018}).  Although the cosmological interpretation of the EDGES result remains controversial (e.g. \citealt{Hills2018arXiv180501421H,Bradley2019ApJ...874..153B,Sims2019MNRAS.488.2904S}), if the signal is indeed proven to be of cosmic origins, X-rays and soft UV radiation from MCGs were likely responsible for its timing (e.g. \citealt{Mirocha2019MNRAS.483.1980M}).

In this work, we explore the imprints of MCGs in the 21-cm signal from the EoR and CD, introducing a distinct population in the public code \cmfast \citep{Mesinger2007ApJ...669..663M,Mesinger2011MNRAS.411..955M},
whose abundance is regulated by an H$_2$-disassociating background (see also \citealt{Fialkov2013MNRAS.432.2909F, Munoz2019PhRvD.100f3538M}).
We build upon the model of \citet{Park2019MNRAS.484..933P}, whose parametrization allows star formation rates (SFRs) to scale non-linearly with the mass of the host halo, thus allowing the source models to be consistent with current UV LF observations (e.g. \citealt{Hassan2016,Mirocha2016,Mutch2016}). We extend this model, allowing MCGs to have their own unique properties, including star formation efficiencies, ionizing escape fractions, and X-ray and soft UV emissivities.
By varying the free parameters in our model, we quantify how the diverse properties of two galaxy populations (atomic and molecular cooling) are imprinted in the global and interferometric 21-cm signals. As a proof of concept, we confront this extended two-population model with the putative EDGES detection, using its timing to constrain the properties of minihalo-hosted galaxies within a Bayesian framework, 21CMMC\footnote{\url{https://github.com/BradGreig/21CMMC}} \citep{Greig2015MNRAS.449.4246G}.  The code developed for this work will be part of the upcoming v3.0.0 release of \cmfast\footnote{\url{https://github.com/21cmfast/21cmFAST}}.

This paper is organized as follows. We present our model in Section \ref{sec:models}.  In Section \ref{sec:xHandTs}, we investigate the impact of the physical properties of our model on the 21-cm signal. In Section \ref{sec:constraint} we perform a Monte Carlo Markov Chain (MCMC) with a subset of model parameters, showing constraints available with and without the EDGES result.  Finally, we conclude in Section \ref{sec:conclusion}. In this work, we use a $\Lambda$CDM cosmology with parameters
$\Omega_{\mathrm{m}}, \Omega_{\mathrm{b}}, \Omega_{\mathrm{\Lambda}}, h, \sigma_8$ and $n_s$ = (0.31, 0.048, 0.69, 0.68, 0.81, and 0.97, consistent with results from the {\it Planck} satellite (e.g. \citealt{Planck2016A&A...594A..13P}).

\section{Modelling galaxies, the IGM, and cosmic radiation fields}\label{sec:models}
\subsection{Star formation and galaxy evolution}\label{subsec:sf_models}

As gas from the IGM accretes onto dark matter halos, its gravitational potential energy is converted into heat.  In order to avoid becoming pressure supported and continue collapsing onto the galaxy at the centre of the halo, gas needs to cool by emitting radiation that must escape the system.  Galaxies can be classified by the dominant cooling channel through which the IGM gas has been accreted onto the halo: (i) atomic-cooling galaxies (ACGs), which predominantly obtained their gas through ${\hone}$ (and $\textsc{H}$e) line transitions efficient at virial temperatures ($T_\mathrm{vir}$) above $10^4 \mathrm{K}$; and (ii) MCGs, in which the gas cools mainly through the $\textsc{H}_2$ rotational--vibrational transitions efficient at $T_\mathrm{vir}{\sim}10^3 {-} 10^4 \mathrm{K}$; most ACGs at high redshifts are ``second-generation'' galaxies, forming out of MCG building blocks. The pre-enrichment by metals as well as the different energetics and cooling processes suggests that the stellar component and interstellar medium (ISM) of ACG and MCG should have different properties. As these properties are currently poorly understood, we describe them through relatively generic and flexible parametric models.  Next, we introduce these for both ACGs and MCGs.

\subsubsection{Atomic-cooling galaxies}\label{subsec:source model}

``Massive'' ACGs ($T_\mathrm{vir}{>}10^5\mathrm{K}$; e.g. \citealt{Kuhlen2012,Mason2015,Liu2016}) have been observed by the {\it Hubble Space Telescope} ({\it HST}).  The resulting non-ionizing UV LFs (e.g. \citealt{Bouwens2015a,Bouwens2016,Finkelstein2015ApJ...810...71F,Livermore2017ApJ...835..113L,Atek2018MNRAS.479.5184A,Oesch2018ApJ...855..105O,Bhatawdekar2019MNRAS.tmp..843B}) give invaluable insight into star formation processes inside these galaxies, ruling out the constant mass-to-light ratio models commonly found in early 21-cm forecasts (e.g. \citealt{Mesinger2011MNRAS.411..955M, Fialkov2013MNRAS.432.2909F}).

Here we build upon the model of \citet[][hereafter \citetalias{Park2019MNRAS.484..933P}]{Park2019MNRAS.484..933P}, which is flexible enough to reproduce observed high-redshift LFs.  This simple model describes the ACG population through power-law scaling relations with the halo mass function (HMF; see also \citealt{Kuhlen2012,Mitra2015,Sun2015,Behroozi2019MNRAS.488.3143B})
Although individual galaxies have much more complicated and stochastic evolution of properties (e.g. \citealt{Mutch2016, Ma2018, Yung2019MNRAS.490.2855Y}), the 21-cm signal on observable scales (${\gsim } 10 \mathrm{Mpc}$) is sourced by $\gsim$ hundreds of galaxies, motivating the use of simple and computationally efficient average scaling relations.

Specifically, we describe the stellar mass of an ACG, $M_*^{\rm atom}$, hosted in a halo with a mass of ${\mvir}$ by
\begin{equation}\label{eq:m*_atom}
M_*^{\rm atom} = \min\left[1, ~f_{*,10}^{\rm atom}\left(\frac{\mvir}{10^{10}\msol}\right)^{\alpha_*}\right] \frac{\Omega_{\mathrm{b}}}{\Omega_{\mathrm{m}}} {\mvir},
\end{equation}
where $f_{*,10}^{\rm atom}$ and ${\alpha_*}$ are the normalization factor and power-law index. More detailed models recover such scaling relations for the bulk of the high-redshift galaxy population (e.g. \citealt{Moster2013MNRAS.428.3121M,Mutch2016,Sun2015,Tacchella2018ApJ...868...92T,Behroozi2019MNRAS.488.3143B,Yung2019MNRAS.490.2855Y}). Note that we do not consider AGN feedback, which is thought to dominate in the most massive galaxies, as these are too rare at high redshifts to shape the 21-cm signal (e.g. \citealt{Mitra2015,Manti2017,Parsa2017,Qin2017a,Ricci2017MNRAS.465.1915R,Garaldi2019MNRAS.483.5301G}).  We also do not include a redshift evolution in this stellar-to-halo mass relation, which is supported by some simulation results (e.g. \citealt{Mutch2016,Xu2016ApJ...833...84X}), although generalizing the model to include a redshift evolution is trivial.

The corresponding average SFR is assumed to be
\begin{equation}\label{eq:sfr}
{\rm SFR}^{\rm atom} = \frac{M_*^{\rm atom}}{t_* H^{-1}\left(z\right)},
\end{equation}
where $t_*$ is a free parameter corresponding to the typical star formation time-scale, defined as a fraction of the Hubble time.  Since the dynamical time of a halo scales with the Hubble time during matter domination, this is analogous to assuming the star formation time scales with the dynamical time.

We include an exponential duty cycle\footnote{{\color{black}The duty cycle is defined as the fraction of halos that harbour star-forming galaxies for a given halo mass. It is used to describe the stochasticity of star formation and can be considered as an occupation fraction (e.g. \citealt{Lippai2009ApJ...701..360L,Miller_2015}).}} to describe the mass function of halos (HMF) that host star-forming ACGs
\begin{equation}\label{eq:hmf_atom}
\phi^{\rm atom} = \frac{\rmd n}{\rmd \mvir} \exp\left({-\dfrac{M_{\rm crit}^{\rm atom}}{\mvir}}\right),
\end{equation}
where ${\rmd n}/{\rmd \mvir}$ is the mass function of all halos.
The exponential term in equation (\ref{eq:hmf_atom}) accounts for inefficient star formation in halos below a characteristic mass scale (i.e. turnover mass)
\begin{equation}\label{eq:m_crit_atom}
M_{\rm crit}^{\rm atom} = \max\left[ M_{\rm crit}^{\rm cool},  M_{\rm crit}^{\rm ion}, M_{\rm crit}^{\rm SN} \right] ~ .
\end{equation}
As can be seen from equation (\ref{eq:m_crit_atom}), we assume that star formation in small ACGs can be limited by three physical processes: (i) inefficient cooling, $M_{\rm crit}^{\rm cool}$; (ii) photoheating feedback from reionization, $M_{\rm crit}^{\rm ion}$; and (iii) supernova feedback, $M_{\rm crit}^{\rm SN}$.

For (i), we assume the ${\hone}$ cooling threshold to be 10$^4$K.  The corresponding halo mass can be expressed as (e.g. \citealt{Barkana2000})
\begin{equation}\label{eq:mcrit_cool}
\frac{M_{\rm crit}^{\rm cool}}{5{\times}10^7 \msol} = \left(\frac{0.678}{h}\right) \left(\frac{0.59}{\mu}\frac{10}{1{+}z}\right)^{1.5}
\left(\frac{\omegam^{\rm z}}{\omegam}\frac{18 {\rm {\rm \pi}}^2}{\Delta_{\rm c}}\right)^{0.5},
\end{equation}
where $\mu$ is the mean molecular weight, $\Delta_c$ is the critical overdensity of halos at collapse in the spherical collapse model, and $\omegam^{\rm z}$ is the matter density  {\color{black}in units of the critical density at redshift $z$}.

For photoheating feedback inside reionized regions of the Universe (ii;  see \citealt{Efstathiou1992MNRAS.256P..43E,Shapiro1994ApJ...427...25S,Thoul1996ApJ...465..608T,Hui1997MNRAS.292...27H};  {\color{black}\citealt{Dijkstra2004ApJ...601..666D}}), we take the functional form from \citet{Sobacchi2014MNRAS.440.1662S}
\begin{equation}\label{eq:mcrit_ion}
\dfrac{M_{\rm crit}^{\rm ion}}{2.8{\times}10^9 \msol}  {=} \left(\frac{f_{\rm bias}\bar{\Gamma}_{\rm ion}}{10^{-12}{\rm s^{-1}}}\right)^{0.17} \left(\frac{10}{1{+}z}\right)^{2.1}\left[1 {-} \left(\frac{1{+}z}{1{+}z^{\rm ion}}\right)^{2}\right]^{2.5},
\end{equation}
where $\bar{\Gamma}_{\rm ion}$ and $z^{\rm ion}$ are the local photoionization rate and the redshift at which the local patch of the IGM was reionized, respectively; the factor $f_{\rm bias}{\approx2}$ accounts for the enhanced photoionization rate at galaxy locations due to their clustering \citep{Mesinger2008MNRAS.390.1071M}. We note that although equation (\ref{eq:mcrit_ion}) is obtained from 1D collapse simulations, it is consistent with
results from more detailed 3D simulations at the relevant redshifts (e.g. \citealt{Noh2014MNRAS.444..503N,Katz2019arXiv190511414K}). We will discuss how to calculate the ionizing background and determine the redshift of ionization in Section \ref{subsubsec:UVionizing}.

Supernova feedback (iii) is probably the least well-understood feedback process.  The dynamic range required to study supernova feedback is enormous.  Thus, its implementation in cosmological simulations is resolution dependent and relies on the choice of subgrid prescription
(e.g. \citealt{DallaVecchia2008,DallaVecchia2012,Hopkins2014MNRAS.445..581H,Keller2014,Hopkins2017}; Gillet et al., in prep.; Pallottini et al., in prep.).  Although it is a free parameter in our model, for this work we assume it to be subdominant compared to inefficient cooling and photoheating in regulating star formation, i.e. taking $M_{\rm crit}^{\rm SN} \le \max\left[M_{\rm crit}^{\rm cool},  M_{\rm crit}^{\rm ion} \right]$  {\color{black}and ignore $M_{\rm crit}^{\rm SN}$ in equation (\ref{eq:m_crit_atom})}. This is a conservative choice in that it maximizes the importance of star formation in small ACGs and minihaloes, which is the focus of this work. Note that supernova feedback is still expected to determine the scaling of the stellar-to-halo mass relation (e.g. \citealt{Moster2013MNRAS.428.3121M,Wyithe2013MNRAS.428.2741W,Dayal2014MNRAS.445.2545D,Mutch2016,Sun2015,Tacchella2018ApJ...868...92T}), even if it is not responsible for a faint end turnover of the LFs.

\subsubsection{Molecular-cooling galaxies}\label{subsec:source model_mcg}
Since star formation can proceed differently in MCGs compared to ACGs, we allow them to have a different stellar-to-halo mass normalization
\begin{equation}\label{eq:m*_mol}
M_*^{\rm mol} = \min\left[1, f_{*,7}^{\rm mol}\left(\frac{\mvir}{10^{7}\msol}\right)^{\alpha_*}\right] \frac{\Omega_{\mathrm{b}}}{\Omega_{\mathrm{m}}} {\mvir} ~ ,
\end{equation}
and calculate their SFRs analogously to equation (\ref{eq:sfr}). We define the mass function of MCG hosts as
\begin{equation}\label{eq:hmf_mol}
\phi^{\rm mol} =\frac{\rmd n}{\rmd \mvir} \exp\left({-\dfrac{M_{\rm crit}^{\rm mol}}{\mvir}}\right)\exp\left({-\dfrac{\mvir}{M_{\rm crit}^{\rm cool}}}\right) ~ .
\end{equation}
The two exponential terms in equation (\ref{eq:hmf_mol}) correspond to duty cycles of halos hosting MCGs, setting both a lower and an upper mass threshold.  The upper mass threshold, $M_{\rm crit}^{\rm cool}$, corresponds to the transition between MCG and ACG, at around $T_\mathrm{vir}\sim10^4{\rm K}$ (see equation \ref{eq:mcrit_cool}). 

It is worth noting that our duty cycles are exponential functions of halo mass, which is a somewhat arbitrary choice.  One impact of this is that the transition from MCGs and ACGs is not a step function at $T_{\rm vir}\sim10^4$ K, as is commonly assumed due to the rapid drop in the atomic cooling curve. However, it is plausible to expect a transition smoother than a step function {\color{black}from the large scatter in the gas temperature-to-halo mass relation (e.g. \citealt{Shang2010MNRAS.402.1249S}). Additionally}, a fraction of galaxies with $T_\mathrm{vir}{>}10^4{\rm K}$ could have obtained most of their gas at earlier times when the H$_2$ cooling channel was dominant.  Similarly, one could have some rare galaxies with ACG-like properties below the cooling threshold, if they occur in pre-enriched dense environments with rapid accretion of cold streams (e.g. \citealt{Qin_2019}).  In practice, these duty cycles serve as window functions over the HMFs to encapsulate two distinct galaxy populations, and our results are not sensitive to the specific choice of window function.

The lower mass threshold for star-forming MCGs is set by cooling and feedback, analogously to equation (\ref{eq:m_crit_atom}) for ACGs
\begin{equation}\label{eq:m_crit_mol}
M_{\rm crit}^{\rm mol} = \max\left[ M_{\rm crit}^{\rm diss},  M_{\rm crit}^{\rm ion}, M_{\rm crit}^{\rm SN} \right] ~ .
\end{equation}
The efficiency of ${\htwo}$ cooling depends on the strength of the dissociating (LW) background, in the energy range 11.2 -- 13.6 eV.  We quantify this using the fitting formulae from {\color{black}\citet{Visbal2015MNRAS.453.4456V}}
\begin{equation}\label{eq:mcrit,diss}
\frac{M_{\rm crit}^{\rm diss}}{2.5\times10^5\msol } = \left(\frac{26}{1{+}z}\right)^{1.5}\left[1+ 22.87\times{J^{\rm 21}_{\rm LW,eff}}^{0.47}\right] ~.
\end{equation}
Here the unitless quantity, $J^{\rm 21}_{\rm LW,eff}$, represents the (local) LW intensity impinging on the MCG
\begin{equation}\label{eq:fshiled}
J^{\rm 21}_{\rm LW,eff} = \frac{J_{\rm LW}}{{\rm 10^{{-}21}{\rm erg\ s^{{-}1}\ Hz^{-1}\ cm^{{-}2}\ sr^{{-}1}}}} \left(1-f_{\rm H_2}^{\rm shield}\right)
\end{equation}
with $J_{\rm LW}$ corresponding to the local (inhomogeneous) LW background (LWB; discussed in Section \ref{subsubsec:LW}), and the factor $f_{\rm H_2}^{\rm shield}$ accounting for self-shielding of star-forming regions by the ISM and the circumgalactic medium of the host galaxy\footnote{Since the column density ratio between $\hone$ and $\htwo$ ($N_{\hone}/N_{\htwo}$) at high redshift is poorly understood (e.g. \citealt{Cen2003ApJ...591...12C}), and self-shielding also depends on the temperature and velocity structure of the ISM (e.g. \citealt{Wolcott-Green2011MNRAS.418..838W}), here we allow $f_{\rm H_2}^{\rm shield}$ to be a free parameter instead of relating it to the typical $\hone$ column density.} (e.g. \citealt{Draine1996ApJ...468..269D,Wolcott-Green2011MNRAS.418..838W}).

In this work, we do not account for a possible additional suppression of star formation in minihaloes due to the relative velocities of dark matter and baryons, imprinted at recombination \citep{Tseliakhovich2010PhRvD..82h3520T}.
The root-mean-square velocity offset at $z\sim20$ is roughly $\sigma_{\rm vb} \sim 0.5{\rm km\ s^{-1}}$ \citep{Munoz2018Natur.557..684M}, which is smaller than the typical circular velocity of minihaloes, $v_{\rm circ} \sim 4{\rm km\ s^{-1}}$.
Therefore, relative velocities are unlikely to be the main bottleneck in feeding gas to MCGs at observable redshifts (e.g.  \citealt{Fialkov2012MNRAS.424.1335F}).  Nevertheless, they do somewhat suppress their cold gas reservoir (e.g. \citealt{Dalal:2010yt,Greif2011ApJ...736..147G,Oleary2012ApJ...760....4O,Schauer2019MNRAS.484.3510S}), which can in turn suppress their SFRs. Although modest, such a decrease in SFRs is correlated on fairly large scales, set by acoustic oscillations prior to recombination.  This might spatially modulate the 21-cm signal in a way that could be detectable with next-generation interferometers (e.g. \citealt{Fialkov2012MNRAS.424.1335F,Munoz2019PhRvD.100f3538M}), providing a standard ruler at CD \citep{Munoz:2019fkt}. We postpone a detailed investigation of this claim to future work (Mu\~noz et al. in prep.).

\subsubsection{UV LFs}\label{subsubsec:UVLF}

\begin{figure*}
	\begin{minipage}{\textwidth}
		\begin{center}
			\includegraphics[width=0.54\textwidth]{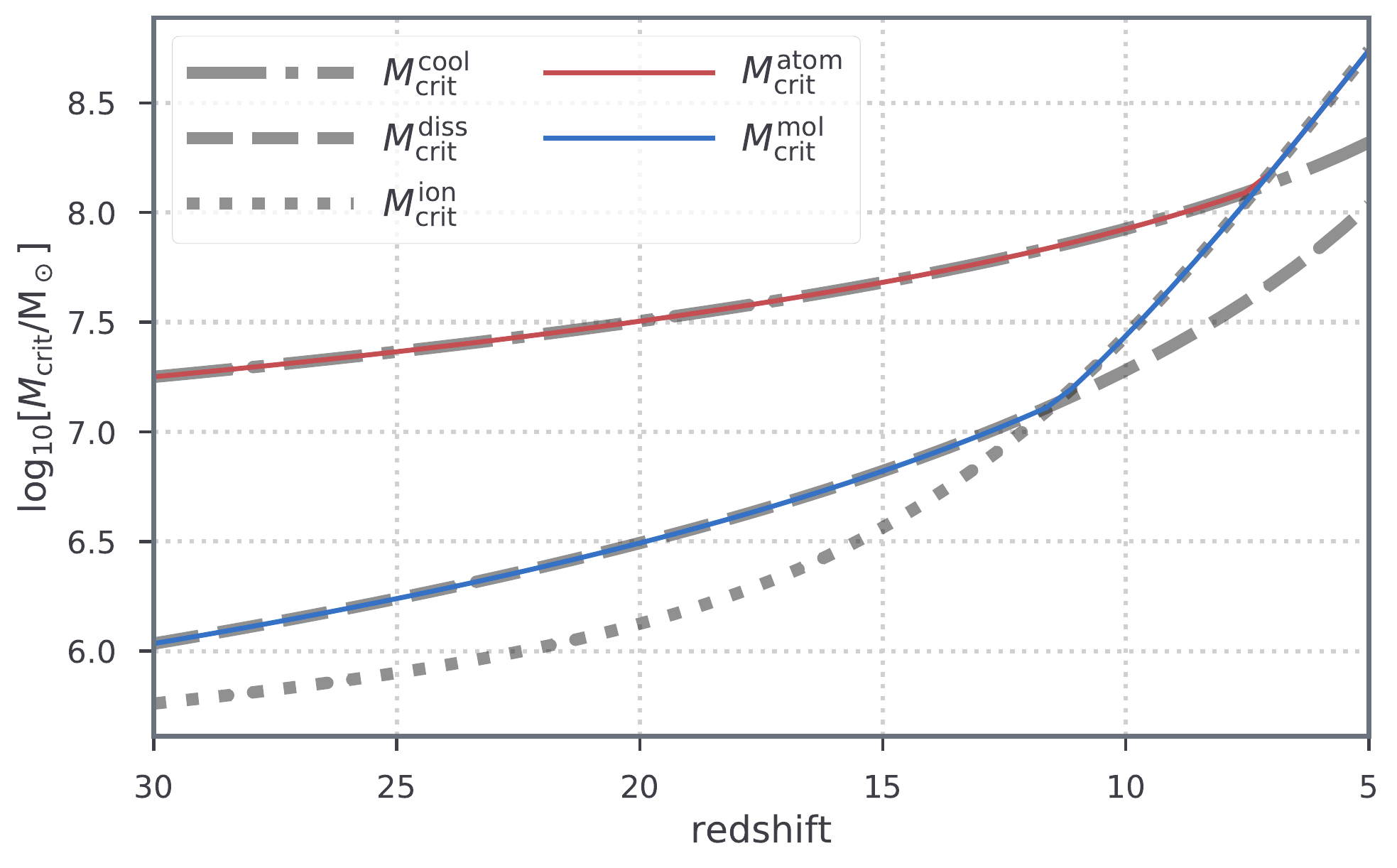}\\\vspace*{-2mm}
			\includegraphics[width=0.92\textwidth]{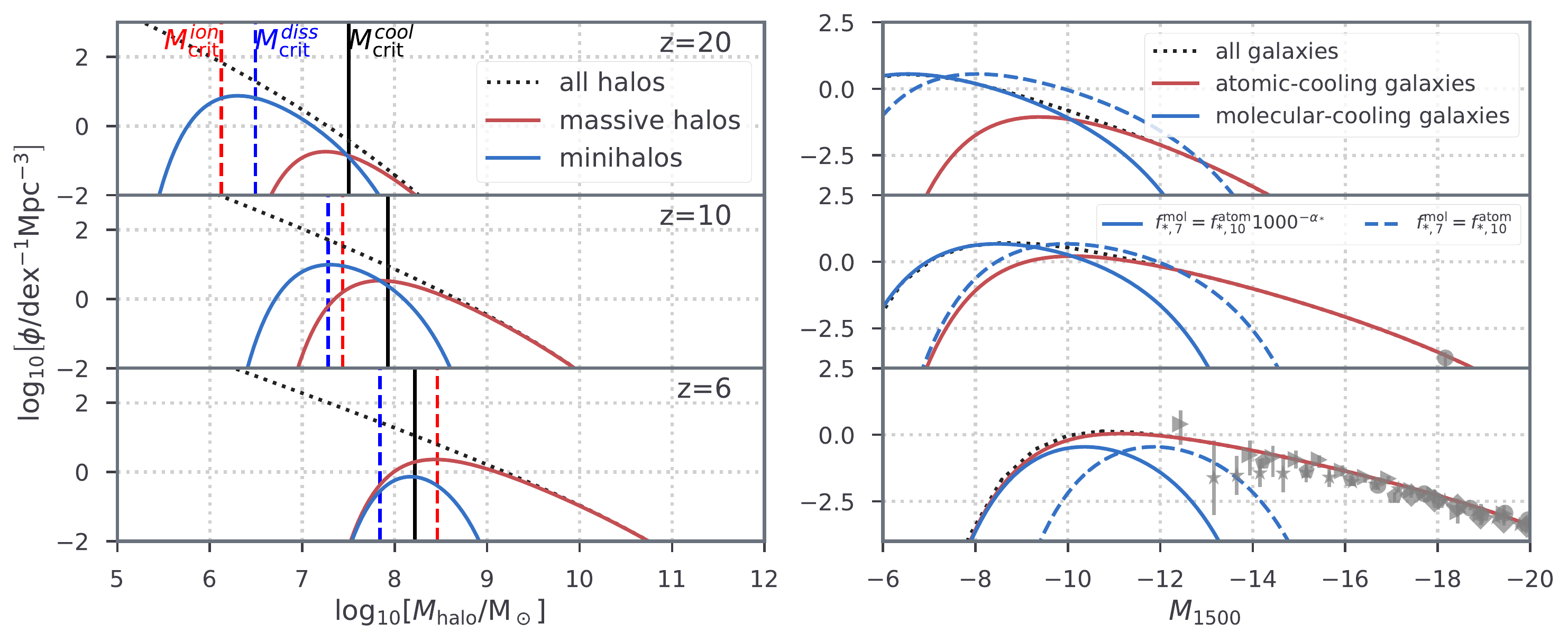}\vspace*{-4mm}
		\end{center}
		\caption{\label{fig:example}An illustration of our two-population source model. \textit{Upper panel:} the evolution of critical masses determined by:  (i)  the atomic cooling efficiency, $M_{\rm crit}^{\rm cool}$; (ii) photodissociation of $\htwo$, $M_{\rm crit}^{\rm diss}$; and (iii) photoheating from reionization, $M_{\rm crit}^{\rm ion}$.  Effects (i) {--} (iii) determine the halo mass scale below which star formation in ACGs ($M_{\rm crit}^{\rm atom}$) and MCGs becomes inefficient ($M_{\rm crit}^{\rm mol}$). \textit{Lower left panel:} mass functions of all halos as well as halos hosting ACGs and MCGs at three typical epochs. The relevant critical masses are marked by vertical lines. \textit{Lower right panel:} corresponding galaxy UV LFs assuming that the stellar-to-halo mass relation of MCGs follows ACGs ($f_{*,7}^{\rm mol}={{1000^{-\alpha_*}}f_{*,10}^{\rm atom}}$). The LF of MCGs with an enhanced star formation efficiency ($f_{*,7}^{\rm mol}=f_{*,10}^{\rm atom}$) and observations  {\color{black}(diamonds: \citealt{Finkelstein2015ApJ...810...71F}; circles: \citealt{Bouwens2015a}; triangles: \citealt{Livermore2017ApJ...835..113L}; stars: \citealt{Atek2018MNRAS.479.5184A}; pentagons: \citealt{Bhatawdekar2019MNRAS.tmp..843B})} are shown for comparison. An ultra-deep {\it JWST} survey could push these measurements $\sim$1--2 mag deeper, making it unlikely to study many MCGs with direct observations.}
	\end{minipage}
\end{figure*}

In order to compare our models with observed LFs from {\it HST}, we convert the SFR to a corresponding intrinsic UV luminosity at $1500\rm{\AA}$ via ${L_{\rm 1500}}/{{\rm SFR}} = 8.7\times10^{27}{\rm erg\ s^{-1} Hz^{-1} {\msol}^{-1} {\rm yr}}$ (e.g. \citealt{Madau2014ARA&A..52..415M}).
This conversion factor can vary by up to a factor of $\sim$2, depending on the IMF, {\color{black}metallicity}, and recent star formation history (e.g. \citealt{Tumlinson2000ApJ...528L..65T, Bromm2001ApJ...552..464B, Eldridge2017PASA...34...58E, Wilkins2019arXiv191005220W}). As it is degenerate with the stellar fraction, a misestimate would imply a bias in constraints on $f_\ast$ from LF observations.  For simplicity, here we use the same conversion factor for both MCGs and ACGs; however, MCGs are generally too faint to be constrained by LF observations (see the bottom right panel of Figure \ref{fig:example}) making our results insensitive to this choice.  In future work, we will expand on this conversion, including the relevant uncertainties in our forward modelling.

\subsubsection{An illustration of our two-population source model}

We illustrate the updated two-population source model in Fig. \ref{fig:example}, assuming MCGs and ACGs follow the same star formation efficiency -- halo mass relation ($f_{*,7}^{\rm mol}={{1000^{-\alpha_*}}f_{*,10}^{\rm atom}}$).  In the top panel, we show the evolution of critical masses determined by:  (i)  the atomic cooling efficiency, $M_{\rm crit}^{\rm cool}$; (ii) photodissociation of $\htwo$, $M_{\rm crit}^{\rm diss}$; and (iii) photoheating from reionization, $M_{\rm crit}^{\rm ion}$.  Also shown are the corresponding halo mass scales below which star formation in ACGs ($M_{\rm crit}^{\rm atom}$) and MCGs becomes inefficient ($M_{\rm crit}^{\rm mol}$), determined by effects (i) -- (iii).  Note that photoheating only becomes dominant in the advanced stages of reionization (e.g.  \citealt{Mesinger2008MNRAS.390.1071M,Ocvirk2018arXiv181111192O,Katz2019arXiv190511414K}).

We select three representative epochs and show the mass functions of all halos (black dotted line), as well as those hosting ACGs (red solid line) and MCGs (blue solid line) in the lower left panel of Fig. \ref{fig:example}. As expected, star formation in ACGs and MCGs is regulated by inefficient cooling in the very early universe. As the intensity of the LWB increases with time, it becomes increasingly difficult for gas to cool {\color{black}in} minihaloes, as denoted by the shift of $M_{\rm crit}^{\rm diss}$ towards higher masses (c.f. equation \ref{eq:mcrit,diss}).
At early times before the bulk of the EoR, the cosmic ${\hii}$ regions are still confined to be proximate to the nascent galaxies; therefore photoheating feedback does not affect most of the volume. Moreover, since the gas responds to the radiation background on roughly the sound-crossing time-scale, photoheating feedback only becomes the dominant negative feedback for both galaxy populations towards the end of the EoR (see the late rise in the dotted gray curve in the top panel).

In the lower right panel of Fig. \ref{fig:example}, we show the corresponding 1500\AA\ UV LFs together with an extreme model in which the star formation efficiency of MCGs is increased by a factor of $1000^{\alpha_*}$ (i.e. $f_{*,7}^{\rm mol}=f_{*,10}^{\rm atom}$; {\it dashed blue curves}). We see that MCGs only dominate the UV LFs at magnitudes fainter than $M_{1500} {\sim}{-}10$ and redshifts higher than $z{\sim}10$.  Thus direct observations of individual MCGs are unlikely even with {\it JWST}, which can extend current {\it HST} observations by $\sim$1--2 mag (i.e. $M_{1500}{\sim}-13$ at $z\sim6$; \citealt{Finkelstein2016PASA...33...37F}; R. Bouwens and P. Oesch, private communication).

Fig. \ref{fig:example} demonstrates the (parametrized) impact of various feedback mechanisms on star formation in MCGs and ACGs.  In the next sections, we describe the calculation of IGM properties as well as the ionizing, LW, X-ray and {\lya} radiation fields -- the essential ingredients that govern these feedback scales and regulate the gas properties responsible for the 21-cm signal.

\subsection{IGM evolution}\label{subsec:igm}

The IGM density and velocity fields are computed at the desired redshift by evolving an initial Gaussian realization with second-order Lagrangian perturbation theory (e.g. \citealt{Scoccimarro1998MNRAS.299.1097S}). The ionization field of the IGM is assumed to be bi-modal -- due to the short mean free path of UV ionizing photons in the neutral IGM and the long average recombination time-scale in the ionized IGM, (almost) fully ionized regions begin appearing and spreading into (almost) fully neutral regions\footnote{As described below, we also account for partial ionization by X-rays, which blurs this distinction for extreme models (e.g. \citealt{RO04, Mesinger2013MNRAS.431..621M}).} (e.g. \citealt{Trac2011ASL.....4..228T,Zahn2011MNRAS.414..727Z}).

Ionized regions of the IGM are identified using the excursion-set procedure described in Section \ref{subsubsec:UVionizing}.  Inside these cosmic HII regions, the temperature is assumed to be ${\sim}10^4$K, while a small amount of residual $\hone$ remains according to photoionization equilibrium with the local (inhomogeneous) UV background (see Section \ref{subsubsec:UVionizing}).

Outside of the cosmic HII regions, the neutral IGM is still impacted by X-ray photons from galaxies {\color{black}\citep{Mesinger2013MNRAS.431..621M}}, which have long mean free paths.  In the neutral IGM, the temperature, $T_{\rm g}$, and ionized fraction, $x_e$, of the gas are evolved from initial conditions computed with \textsc{recfast} \citep{Seager1999ApJ...523L...1S}, according to the following differential equations:
\begin{equation}\label{eq:x_e}                 
\dot{x}_e = {-} \alpha_{\rm A} C_{\rm sub} x_e^2 n_{\rm b}f_{\rm H} + \Lambda_{X} 
\end{equation}
with $\alpha_{\rm A}$, $C_{\rm sub}$, $n_{\rm b}$, $f_{\rm H}$ and $\Lambda_{X}$ representing the case-A recombination coefficient, subgrid clumping factor, number density of baryons in the simulation cell, number fraction of hydrogen, and the X-ray ionization rate per baryon, respectively, and
\begin{equation}\label{eq:T_g}
\frac{3}{2} \left(1{+}x_e\right) \dot{T}_{\rm g}= \left(1{+}x_e\right) \frac{\dot{n}_{\rm b}}{n_{\rm b}} {T}_{\rm g} {-} \frac{3}{2} {\dot{x}_e} {T}_{\rm g}{+}  k_{\rm B}^{-1}\left(\varepsilon_{X}  {+}   \varepsilon_{\rm CMB}\right)
\end{equation}
with $k_{\rm B}$ being the Boltzmann constant, $\varepsilon_{X}$ and $\varepsilon_{\rm CMB}$ (in units of ${\rm erg\ s^{-1}}$) correspond to the heating rate per baryon due to X-rays and CMB photons, respectively. Note that the terms on the right side of equation (\ref{eq:x_e}) refer to recombinations and ionization with X-rays while those of equation (\ref{eq:T_g}) correspond to heating/cooling due to structure formation, changing species, X-ray and Compton heating \citep{Seager1999ApJ...523L...1S}, respectively. We ignore other heating processes that are expected to be subdominant at the relevant redshifts, such as dark matter annihilation or shock heating (e.g. \citealt{Furlanetto2006PhR...433..181F,McQuinn2012ApJ...760....3M,Evoli2014JCAP...11..024E,Lopez-Honorez2016JCAP...08..004L}).  We describe the calculation of X-ray ionization and heating rates in Section \ref{subsubsec:xraysandlya}.

\subsection{Radiation fields}\label{subsec:photon model}
Cosmic radiation fields regulate the ionization and thermal state of the IGM, as well as the star formation feedback processes described previously.  Here we summarize how we calculate the ionizing, LW,  X-ray and {\lya} radiation fields.

\subsubsection{UV ionizing photons}\label{subsubsec:UVionizing}
We follow an excursion-set approach \citep{Furlanetto2004ApJ...613....1F} to identify cosmic ${\hii}$ regions -- counting the number of ionizing photons in spheres of decreasing radius around each IGM parcel. A cell, centred at $({\bf r}, z)$, is considered ionized if, at any radius $R$, 
\begin{equation}\label{eq:ionization}
\bar{n}_{\rm ion} \ge \left(1+ \bar{n}_{\rm rec}\right)\left(1 - \bar{x}_e\right).
\end{equation}
\noindent Here, $\bar{n}_{\rm ion}$ is the cumulative number of ionizing photons per baryon, $\bar{n}_{\rm rec}$ is the cumulative number of recombinations per baryon, and $\bar{x}_e$ accounts for X-ray ionizations as described in the previous section. The averaging is performed over the spherical region with a radius of $R$ and a corresponding overdensity of $\delta_{\rm R|_{{\bf r},z}} = \rho_{\rm b} / \bar{\rho}_{\rm b} - 1$.

The left-hand side of equation (\ref{eq:ionization}) is calculated using an updated form from equation (15) of \citetalias{Park2019MNRAS.484..933P}, accounting for both galaxy populations (i.e. MCG and ACG; $i \in \left\{{\rm mol}, {\rm atom}\right\}$). Specifically, the cumulative number of ionizing photons per baryon in a spherical IGM patch is
\begin{equation}\label{eq:dotn_ion}
\bar{n}_{\rm ion}\left({\bf r},z|R,\delta_{\rm R|_{{\bf r},z}}\right) {=}  \rho_{\rm b}^{-1}\sum_{i\in\left\{\substack{{\rm mol,}\\{\rm atom}}\right\}} \int {\rm d}M_{\rm vir} {\phi}^{i} M_*^{i} n_{\gamma}^{i} f_{\rm esc}^{i}.
\end{equation}
In this equation\footnote{The cumulative photon number density, $\bar{n}_{\rm ion}$, is computed via trapezoidal integration over redshift snapshots in each region.  Our approximate treatment of photoheating feedback has a somewhat too rapid evolution at the final stages of reionization (e.g. \citealt{Noh2014MNRAS.444..503N,Katz2019arXiv190511414K}). To compensate for this, we compute the ionizing photon number assuming the same the critical mass threshold (i.e. $M_{\rm crit}^{\rm atom}$ and {\color{black}$M_{\rm crit}^{\rm mol}$}) between two consecutive snapshots, which also ensures $\bar{n}_{\rm ion}$ to increase monotonically with time.}
\begin{enumerate}
	\item[1.]${\phi}^{i} \left(M_{\rm vir}, {\bf r}, z | R, \delta_{\rm R|_{{\bf r},z}} \right)$ represents the differential number density of halos of mass $M_{\rm vir}$ that host ACGs or MCGs (see equations \ref{eq:hmf_atom} and \ref{eq:hmf_mol}), in a spherical volume centred at $({\bf r}, z)$ of radius $R$ and overdensity $\delta_{\rm R|_{{\bf r},z}}$, computed using the hybrid conditional mass function suggested by \citet{Barkana2005ApJ...626....1B}, and adapted to quasi-linear density fields in \citet{Mesinger2011MNRAS.411..955M};
	\item[2.]$M_*^{i}$ refers to the stellar mass of ACGs and MCGs following equations (\ref{eq:m*_atom}) and (\ref{eq:m*_mol});
	\item[3.]$n_{\gamma}^{i}$ corresponds to the number of ionizing photons emitted per stellar baryon. Following \citetalias{Park2019MNRAS.484..933P}, $n_{\gamma}^{\rm atom}=5\times10^3$ is chosen for ACGs. We note that, similarly to the SFR-$L_{\rm 1500}$ conversion factor (see Section \ref{subsubsec:UVLF}), $n_{\gamma}$ depends on the IMF and this value is close to a PopII star-dominated galaxy assuming a Salpeter IMF. We instead choose $n_{\gamma}^{\rm mol} = 5\times10^4$ for MCGs since they should preferentially host metal-free, PopIII stars, expected to have a higher ionizing photon emissivity (e.g. \citealt{Tumlinson2000ApJ...528L..65T,Schaerer2002A&A...382...28S,Bromm2004ARA&A..42...79B}). It is worth noting that the large degeneracy between the efficiency of ionizing photon production and the ionizing escape fraction (see below) means that uncertainties in the former (which we hold fixed in this work) can be subsumed in the inferred constraints on the latter.
	\item[4.] $f_{\rm esc}^{i}$ is the escape fraction defined as the number ratio of ionizing photons that reach the IGM to those intrinsically emitted. It is determined by the ISM properties, such as the ${\hone}$ filling factor, dust, and their distribution with respect to star formation sites. In low-mass halos, the gravitational potential is shallow, facilitating the creation of low column density channels through which ionizing photons can escape. This is expected to result in a negative correlation between $f_{\rm esc}$ and the host halo mass, $M_{\rm vir}$ (e.g. \citealt{Ferrara2013MNRAS.431.2826F,Kimm2014,Paardekooper2015MNRAS.451.2544P,Xu2016ApJ...833...84X}, but also see \citealt{Ma2015MNRAS.453..960M,Naidu2019arXiv190713130N}). We adopt a power-law relation for the escape fraction to halo mass, allowing both the normalization and scaling to be different between MCGs and ACGs (see e.g. fig. 15 in \citealt{Xu2016ApJ...833...84X})
\begin{equation}\label{eq:f_esc}
	f_{\rm esc}^{{\rm atom}({\rm mol})} =\min\left[1, {\color{black}f_{\rm esc,{{10}({7})}}^{{\rm atom}({\rm mol})}} \left(\frac{M_{\rm vir}}{10^{{10}({7})}{\msol}}\right)^{\alpha_{\rm esc}^{{\rm atom}({\rm mol})}}\right].
	\end{equation}
\end{enumerate}

To account for inhomogeneous recombinations, we follow \citet{Sobacchi2014MNRAS.440.1662S} and calculate the number of recombinations per baryon by
\begin{equation}\label{eq:nrec}
\begin{split}
n_{\rm rec}\left({\bf r},z \right) {=} \int_{z_{\rm ion}}^{z} {\rm d z^\prime}\frac{\rm d t}{\rm d z^\prime} \int_{0}^{18{\rm \pi}^2} {\rm d}\Delta_{\rm sub} \frac{\rm d n}{\rm d \Delta_{\rm sub}}\times \\
 \alpha_{\rm B} \bar{n}_{\rm b} f_{\rm H} \Delta_{\rm cell}^{-1}\Delta_{\rm sub}^2 \left(1{-}x_{\rm \hone, sub}\right)^2,
\end{split}
\end{equation}
where $z_{\rm ion}({\bf r})$ is the reionization redshift of the cell; the upper limit of integration, $18{\rm \pi}^2$, corresponds to the overdensities of halos in the spherical collapse model; ${\rm d n}/{\rm d \Delta_{\rm sub}}\left(z^\prime, \Delta_{\rm sub}| \Delta_{\rm cell}\right)$ is the probability distribution function (PDF) of the subgrid (unresolved) overdensities, $\Delta_{\rm sub}$, taken from \citet{Miralda2000ApJ...530....1M} and adjusted for the mean overdensity of the cell, $\Delta_{\rm cell}\equiv n_{\rm b}/\bar{n}_{\rm b}$, according to \citet{Sobacchi2014MNRAS.440.1662S}; $\alpha_{\rm B}$ is the case-B recombination coefficient evaluated at $T_g = 10^4{\rm K}$; and the fraction of residual neutral hydrogen inside the ionized region, $x_{\hone, {\rm sub}}\left(z^{\prime}, \Delta_{\rm cell}, T_{\rm g}, \bar{\Gamma}_{\rm ion}\right)$, is evaluated assuming photoionization equilibrium and accounting for attenuation of the local ionizing background according to \citet{Rahmati2013MNRAS.430.2427R}.

Inside each cosmic {$\hii$} region, we compute the local, average photoionization rate following \citet{Sobacchi2014MNRAS.440.1662S}
\begin{equation}\label{eq:gamma}
{\bar{\Gamma}_{\rm ion}\left({\bf r},z\right)} = \left(1+z\right)^2 R\sigma_{\rm H}\frac{\alpha_{\rm UVB}}{\alpha_{\rm UVB}+\beta_{\rm H}}\bar{n}_{\rm b}{\dot{\bar{n}}_{\rm ion}},
\end{equation}
where $\alpha_{\rm UVB}$ corresponds to the UVB spectral index, $\beta_{\rm H} \approx 2.75$ is the HI photoionization cross-section spectral index, $R$ is the local mean free path -- approximated by the largest radius at which equation (\ref{eq:ionization}) is satisfied,
and $\dot{\bar{n}}_{\rm ion}$ represents the ionizing photon production rate following equation (\ref{eq:dotn_ion}) with $M_*$ being replaced by the SFR (see equation \ref{eq:sfr}).   The ionizing background inside cosmic HII regions is used to calculate the critical mass below which photoheating quenches star formation (c.f. equation \ref{eq:mcrit_ion}), as well as for computing subgrid recombinations (c.f. equation \ref{eq:nrec}).

\subsubsection{LW photons}\label{subsubsec:LW}

Compared to ionizing photons, LW photons have much longer mean free paths in the high-redshift Universe. Therefore, to calculate the LW radiation field at $z$, we must account for distant galaxies, integrating back along the light-cone to include galaxies at higher redshifts, $z^\prime\ge z$, and redshifting the emitted spectrum, $\nu^\prime = \nu \frac{1+z\prime}{1+z}$.

The large resonant cross-section in the Lyman series $\nu_{\rm n}\equiv \nu_{\rm H}\left(1-n^{-2}\right)$ with $\nu_{\rm H}{=}3.29\times10^{15}{\rm GHz}$ being the Lyman limit frequency and $n\in\left[2, 23\right]$\footnote{Following \citet{Barkana2005ApJ...626....1B}, higher order (${>}23$) Lyman transitions are ignored as their contribution is negligible.} provides a barrier for LW photons -- setting a maximum redshift, $z_{{\rm max}}$, from which they can reach $z$
\begin{equation}
\frac{1 + z_{{\rm max}}\left(n\right)}{1+z} =  \frac{1-(n+1)^{-2}}{1-n^{-2}}.
\end{equation}
Equivalently, there is a highest order of Lyman transition, $n_{\rm max}\left(z\right)$, for a given redshift ($z^\prime$), above which photons will redshift into the $n$th-order Lyman transition and be absorbed in the IGM. {\color{black}Note that absorptions of the LW radiation due to the presence of $\htwo$ in the IGM are not considered in this work \citep{Haiman2000ApJ...534...11H,Ricotti2001ApJ...560..580R}}. 
\begin{figure}
	\begin{minipage}{\columnwidth}
		\begin{center}
			\includegraphics[width=\textwidth]{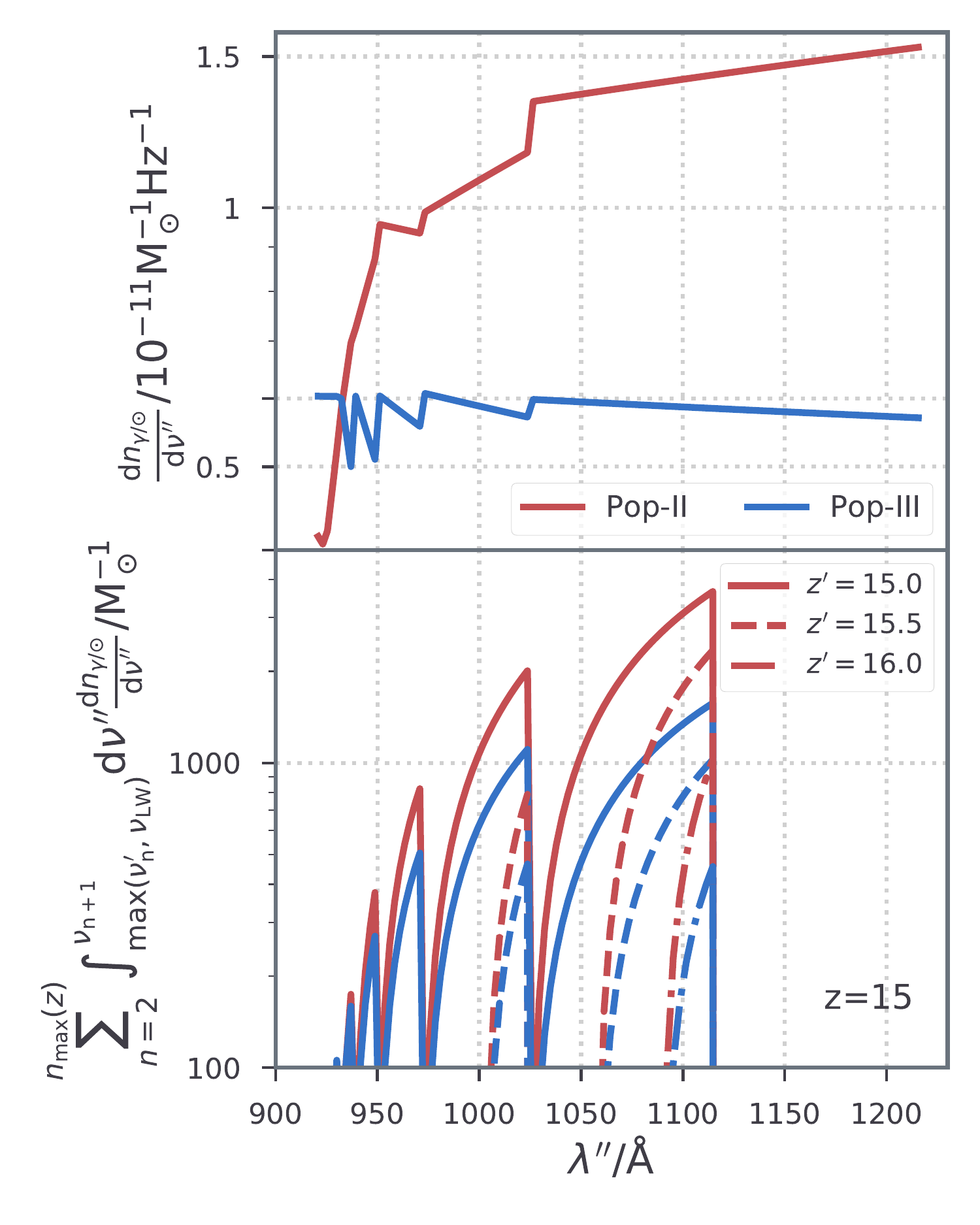}\vspace*{-5mm}
		\end{center}
		\caption{\label{fig:spectra}\textit{Top panel:} PopII- and PopIII-dominated stellar spectra \citep{Barkana2005ApJ...626....1B} used in this work for ACGs and MCGs, respectively. \textit{Bottom panel:} normalized emissivity of the LWB at $z=15$. Photons from higher redshifts ($z^\prime=15.5$ and 16) that contribute to the $z=15$ background are shown in the received frame (i.e. $z=15$) for comparison -- for a given $z^\prime$, there is a maximum energy level in the Lyman series above which photons will be absorbed in the IGM before reaching $z$.}
	\end{minipage}
\end{figure}
Similarly to the {\lya} background calculation of direct stellar emission in
\citet{Mesinger2011MNRAS.411..955M}, the LWB is evaluated with a sum over the Lyman series
(see also e.g.  \citealt{Pritchard2007MNRAS.376.1680P,Ahn2009ApJ...695.1430A,Fialkov2013MNRAS.432.2909F,Munoz2019PhRvD.100f3538M}). After rearranging the integral over redshift and the sum over Lyman series, we obtain the LW radiation intensity, $J_{\rm LW}$ in units of ${\rm{\rm erg\ s^{-1}Hz^{-1}cm^{-2}sr^{-1}}}$, by
\begin{equation}\label{eq:J_LW}
J_{\rm LW} \left({\bf r},z | R, \delta_{\rm R|_{{\bf r},z}} \right) =\frac{\left(1+z\right)^3}{\rm 4{\rm \pi}} \int_{z}^{\infty} {\rmd}z^\prime\frac{{c \rmd}t}{{\rmd}z^\prime} {\epsilon}_{\rm LW}e^{-\tau_{\rm LW}},
\end{equation}
where we assume that the LW photons are only attenuated at resonance, and the corresponding emissivity becomes
\begin{equation}\label{eq:e_LW}
\begin{split}
{\epsilon}_{\rm LW}\left({\bf r},z^\prime|z,R,\delta_{\rm R|_{{\bf r},z}}\right)& =  \sum_{i\in\left\{\substack{{\rm mol,}\\{\rm atom}}\right\}}  \int {\rm d}M_{\rm vir} {\phi}^{i} {\rm SFR}^i  \times \\
& \sum_{n=2}^{n_{\rm max}(z)}  \int_{\max\left(\nu_{\rm n}^\prime, \nu_{\rm LW}\right)}^{\nu_{{\rm n+1}}}\frac{{\rmd}n_{\gamma/\odot}^{i}}{{\rmd}\nu^{\prime\prime}} h {\rmd}\nu^{\prime\prime}.
\end{split}
\end{equation}
When computing the LW emissivity\footnote{When estimating the radiation background of LW (as well as X-ray and {\lya}), we assume that $M_{\rm crit}^{\rm ion}$ is less than $M_{\rm crit}^{\rm diss}$ for the sake of computational efficiency.
This is a valid assumption for the very high redshifts at which the MCG contribution is non-negligible, since photoheating feedback is only dominant after the bulk of reionization (see Fig. 1 and associated discussion).}, we use the PopII- and PopIII-dominated spectral energy distributions (SEDs), ${{\rmd}n_{\gamma/\odot}^{i}}/{{\rmd}\nu^{\prime\prime}}$ (number of photons per mass in stars per unit frequency), from \citet{Barkana2005ApJ...626....1B} for ACGs and MCGs, respectively.  These are shown in the top panel of Fig. \ref{fig:spectra}. They follow piece-wise power laws between pairs of $\nu_{\rm n}$ and $\nu_{\rm n+1}$ with normalizations and scaling indices chosen to reproduce results from stellar-population synthesis models \citep{Leitherer1999ApJS..123....3L,Bromm2001ApJ...552..464B}. We also present the integral in equation (\ref{eq:e_LW}) with the current and higher redshifts being $z=15$ and $z^\prime=15.5$ and 16 in the lower panel.  We see that only a fraction of high-redshift photons between several low-order Lyman transitions can make a contribution to the radiation background of LW at lower redshifts because  of resonant scattering -- the so-called ``picket fence'' absorption (e.g.  \citealt{Haiman1997ApJ...476..458H, Ahn2009ApJ...695.1430A,Fialkov2013MNRAS.432.2909F}).
We then use equations (\ref{eq:mcrit,diss}) and (\ref{eq:fshiled}) to calculate the corresponding LW feedback on MCG star formation.

\subsubsection{X-rays and {\lya} photons}\label{subsubsec:xraysandlya}
We extend \citet{Mesinger2011MNRAS.411..955M} and \citetalias{Park2019MNRAS.484..933P} when estimating the X-ray heating and {\color{black}ionization as well as} {\lya} coupling to allow for both galaxy populations.  We give a brief review of the relevant calculation and encourage readers to follow these two papers for more details.

We start with an assumption that the X-ray emission from all galaxies  (MCG and ACG; $i \in \left\{{\rm mol}, {\rm atom}\right\}$) follows a power law with an energy index of $\alpha_{\rm X}$ and a specific luminosity of
\begin{equation}\label{eq:xspec}
\frac{{\rm d}L_{\rm X/\dot{\odot}}^{i}}{{\rm d}E}\Big(E\Big) {=} L_{\rm X<2keV/\dot{\odot}}^{i} \left(\int_{E_0}^{\rm 2keV}{\rm d}EE^{-\alpha_{\rm X}}\right)^{-1}E^{-\alpha_{\rm X}},
\end{equation}
where $E_0$ represents the minimum energy that an X-ray photon needs to escape from the host galaxy into the IGM [for reference, \citet{Das_2017} estimate $E_0 \sim$ 0.5 keV] while $L_{\rm X<2keV/\dot{\odot}}^{i}$ is the total luminosity between $E_0$ and 2keV.

At these redshifts, the dominant source of soft X-rays (which is relevant for heating/ionizing the IGM) are expected to be High-Mass X-ray Binaries (HMXBs) and potentially also the hot ISM (e.g. \citealt{Fragos2013ApJ...764...41F,Sanderbeck_2018}). Both of these have luminosities that scale with the SFR of the host galaxy (e.g. \citealt{Mineo2012,Fragos2013ApJ...764...41F,Pacucci2014}).  Thus, the ``$/\dot{\odot}$'' in equation (\ref{eq:xspec}) indicates the quantity is per unit SFR -- e.g. $L_{\rm X<2keV/\dot{\odot}}^{\rm mol}$ and $L_{\rm X<2keV/\dot{\odot}}^{\rm atom}$ represent the soft-band X-ray luminosities per SFR for MCGs and ACGs, respectively, which are considered free parameters in our model. Next, we link the X-ray radiation intensity, $J_X$ in units of ${\rm{\rm erg\ s^{-1}keV^{-1}cm^{-2}sr^{-1}}}$, to star formation following\footnote{For the sake of computing efficiency, we follow \citet{Mesinger2011MNRAS.411..955M} and approximate $e^{-\tau_{\rm X}}{=}0$ when ${\tau_{\rm X}}{\ge}1$ and 1 otherwise.  In practice, this approximation makes virtually no impact on the 21-cm power spectrum evolution (e.g. \citealt{Das_2017}).} equation (\ref{eq:J_LW}) with the emissivity term (i.e. ${\epsilon}_{\rm LW}$) being replaced by
{\color{black}
\begin{equation}\label{eq:e_X}
{\epsilon}_{X}\left({\bf r},z^\prime\right) =  \sum_{i\in\left\{\substack{{\rm mol,}\\{\rm atom}}\right\}}  \int {\rm d}M_{\rm vir} {\phi}^{i} {\rm SFR}^i  \frac{{\rm d}L_{\rm X/\dot{\odot}}^{i}}{{\rm d}E}.
\end{equation}
}
Note that the emissivity is evaluated in the rest frame, $E^\prime = E \left(1+z^\prime\right)/\left(1+z\right)$. The ionization (see equation \ref{eq:x_e}) and heating rates per baryon by X-rays (see equation \ref{eq:T_g}) are then computed as follows
\begin{equation}
\Lambda_{X}\left({\mathbf r}, z\right) =  \int_{E_0}^{\infty} {\rm d}E \frac{4{\rm \pi} J_X}{E} \sum_{j}  x^{j} \sigma^{j}  f^{j} \left[{(E{-}E_{\rm th}^{j})}\sum_{k} \frac{f_{\rm ion}^{k}}{E_{\rm th}^{k}} + 1 \right] 
\end{equation}
and
\begin{equation}\label{eq:xheating}
\varepsilon_{X}\left({\mathbf r}, z\right) = \int_{E_0}^{\infty} {\rm d}E \frac{4{\rm \pi} J_X}{E} \sum_{j}  x^{j} \sigma^{j} f^{j}{(E{-}E_{\rm th}^{j})} f_{\rm heat}
\end{equation}
where $f^{j}$ is the number fraction of each species, $j$, with $j\in\left[\mathrm{H}{\textsc{I}}, \mathrm{He}{\textsc{I}}, \mathrm{He}{\textsc{II}}\right]$,
$\sigma^{j}$ is the ionization cross-section,
and $E_{\rm th}^{j}$ is the corresponding energy; $f_{\rm heat}$ and $f_{\rm ion}^{k}$ represent the fraction of the electron energy after ionization, $E{-}E_{\rm th}^{j}$, that contributes to heating or secondary ionization of each species {\color{black}\citep{Furlanetto2010MNRAS.404.1869F}}; and $x^{j} \equiv 1{-} x_e$ when $j\in\left[\mathrm{H}{\textsc{I}}, \mathrm{He}{\textsc{I}}\right]$ or $x_e$ for $\mathrm{He}{\textsc{II}}$ {\color{black}represents the secondary ionization fractions (see Section \ref{subsec:igm})}.

The {\lya} background component coming from direct stellar emission is computed by integrating the emissivity along the light-cone. The evaluation of this background, $J_{\alpha}^{*}$ in units of ${\rm s^{-1}Hz^{-1}cm^{-2}sr^{-1}}$, follows equation (\ref{eq:J_LW}) with the emissivity term (i.e. ${\epsilon}_{\rm LW}$) being replaced by the effective photon number emissivity
\begin{equation}
{\epsilon}_{\alpha}^{*}\left({\bf r},z^\prime\right)  {=}  \sum_{i\in\left\{\substack{{\rm mol,}\\{\rm atom}}\right\}}  \int {\rm d}M_{\rm vir} {\phi}^{i} {\rm SFR}_*^{i} \sum_{n=2}^{n_{\rm max}\left(z\right)}\frac{{\rmd}n_{\gamma/\dot{\odot}}^{i}}{{\rmd}\nu^{\prime\prime}} f_{\rm recycle}\left(n\right),
\end{equation}
where $f_{\rm recycle}$ is the fraction of absorbed photons at the n-th Lyman resonance level that are re-emitted at {\lya} \citep{Hirata2006MNRAS.367..259H,Pritchard2007MNRAS.376.1680P}.

On the other hand, the {\lya} background due to X-ray excitation (in units of ${\rm s^{-1}Hz^{-1}cm^{-2}sr^{-1}}$) can be linked to the heating rate (see equation \ref{eq:xheating}) following
\begin{equation}
\end{equation}
where $f_{\alpha}$ is the fraction of the electron energy ($E-E_{\rm th}^{j}$) that contributes to emitting {\lya} photons with a frequency of ${\nu_\alpha}\equiv 2.47\times10^{15}{\rm Hz}$. The total {\lya} background that is used to evaluate the {\lya} coupling coefficient (see the following section) is the sum of both X-ray and stellar contribution
\begin{equation}
J_{\alpha,{\rm eff}} \times {\rm s^{-1}Hz^{-1}cm^{-2}sr^{-1}} = \left(J_\alpha^{\rm X} + J_\alpha^*\right) \times S_{\alpha}
\end{equation}
where $S_\alpha$ is a quantum mechanical correction factor of order unity \citep{Hirata2006MNRAS.367..259H}.

\section{Modelling the 21-cm signal}\label{sec:xHandTs}

We compute the inhomogeneous 21-cm brightness temperature according to equation (\ref{eq:dtb_define}), albeit with the subgrid non-linear treatment of redshift space distortions and without assuming the optically thin limit (\citealt{Greig2018MNRAS.477.3217G}; see also \citealt{Datta2012MNRAS.424.1877D,Datta2014MNRAS.442.1491D,Mao2012MNRAS.422..926M,Jensen2013MNRAS.435..460J}). The ionization and density fields were discussed previously. The spin temperature is computed according to
\begin{equation}
\label{eq:Tspin}
T_{\rm s}^{-1} = \frac{T_{\rm CMB}^{-1} + \left(x_\alpha + x_{\rm c}\right) T_{\rm g}^{-1}}{1+x_\alpha + x_{\rm c}},
\end{equation}
with the collisional, $x_{\rm c}$, and {\lya} coupling coefficients \citep{Wouthuysen1952AJ.....57R..31W}, $x_{\alpha}$, being calculated by
\begin{equation}\label{eq:x_c}
{x_{\rm c}}= \left(\frac{T_{\rm CMB}}{0.0628{\rm K}}\right)^{-1} \sum_{i\in\left\{\substack{{\rm e,p,\hone}}\right\}} \frac{n_{\rm b} f^\prime_{i} \kappa_{i}}{2.85\times10^{-15}{\rm s}^{-1}}
\end{equation}
and
\begin{equation}\label{eq:x_a}
x_\alpha = 1.7\times 10^{11} \left(1+z\right)^{-1} J_{\alpha,{\rm eff}},
\end{equation}
where
$f^\prime_{i}$ and $\kappa_{i}$ with $i\in\left[e, p, \hone \right]$ represent the number fractions of free electrons, protons and neutral hydrogen and their cross-sections with $\hone$ taken from \citet{Zygelman2005ApJ...622.1356Z} and \citet{Furlanetto2007MNRAS.374..547F}. The IGM only becomes visible in contrast against the CMB if (at least) one of the coupling coefficients in equation (\ref{eq:Tspin}) is non-negligible. 

\subsection{Building physical intuition -- general trends of the reference model}

\begin{figure*}
	\begin{minipage}{\textwidth}
		\hspace*{-4mm}
		\includegraphics[width=1.05\textwidth]{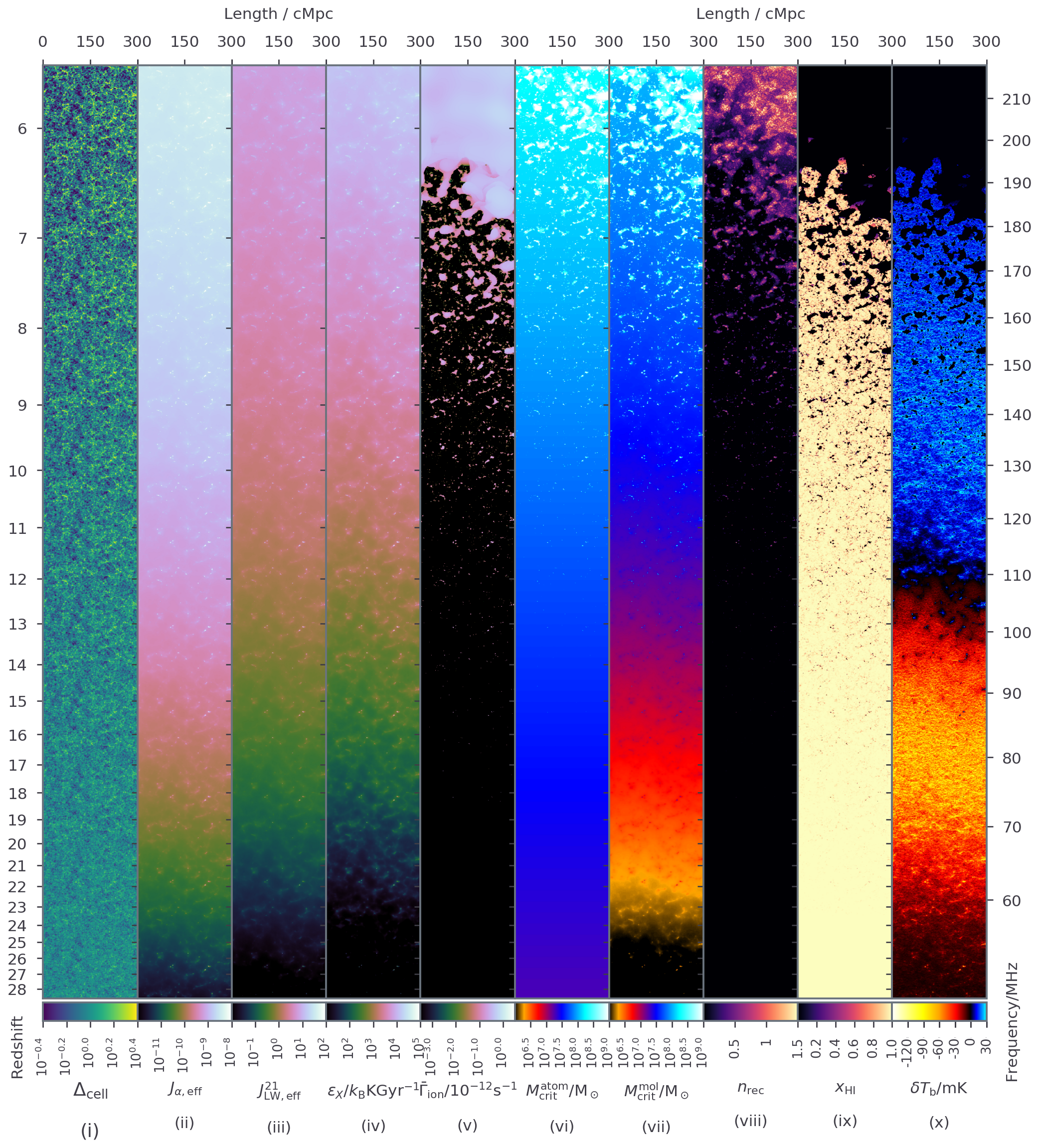}
		\caption{\label{fig:reference}
			Lightcones for the reference model including, from left to right:
			(i) overdensity ($\Delta_{\rm cell}$);
			(ii) {\lya} ($J_{\alpha, {\rm eff}}{\times}\rm s^{-1}Hz^{-1}cm^{-2}sr^{-1}$);
			(iii) LW ($J^{\rm 21}_{\rm LW,eff}{\times}{\rm 10^{{-}21}{\rm erg\ s^{-1}\ Hz^{-1}\ cm^{-2}\ sr^{-1}}}$);
			(iv) X-ray heating ($\varepsilon_{X}$ in units of $k_\mathrm{B} {\rm K Gyr}^{-1}$);
			(v) locally averaged UVB ($\bar{\Gamma}_{\rm ion}$ in units of $10^{-12}\rm{s}^{-1}$);
			(vi) critical halo mass for star formation in ACGs ($M_{\rm crit}^{\rm atom}$/${\msol}$);
			(vii) critical halo mass for star formation in MCGs ($M_{\rm crit}^{\rm mol}$/${\msol}$);
			(viii) cumulative number of recombinations per baryon ($n_{\rm rec}$);
			(ix) neutral hydrogen fraction ($x_{\hone}$); and
			(x) the 21-cm brightness temperature ($\delta T_{\rm b}$ in units of ${\rm mK}$).\newline
			{(A high-resolution version of this figure is available at \url{http://homepage.sns.it/mesinger/Media/light-cones_minihalo.png}.})
		}
	\end{minipage}
\end{figure*}

\begin{figure}
	\begin{minipage}{\columnwidth}
		\begin{center}
			\includegraphics[width=\textwidth]{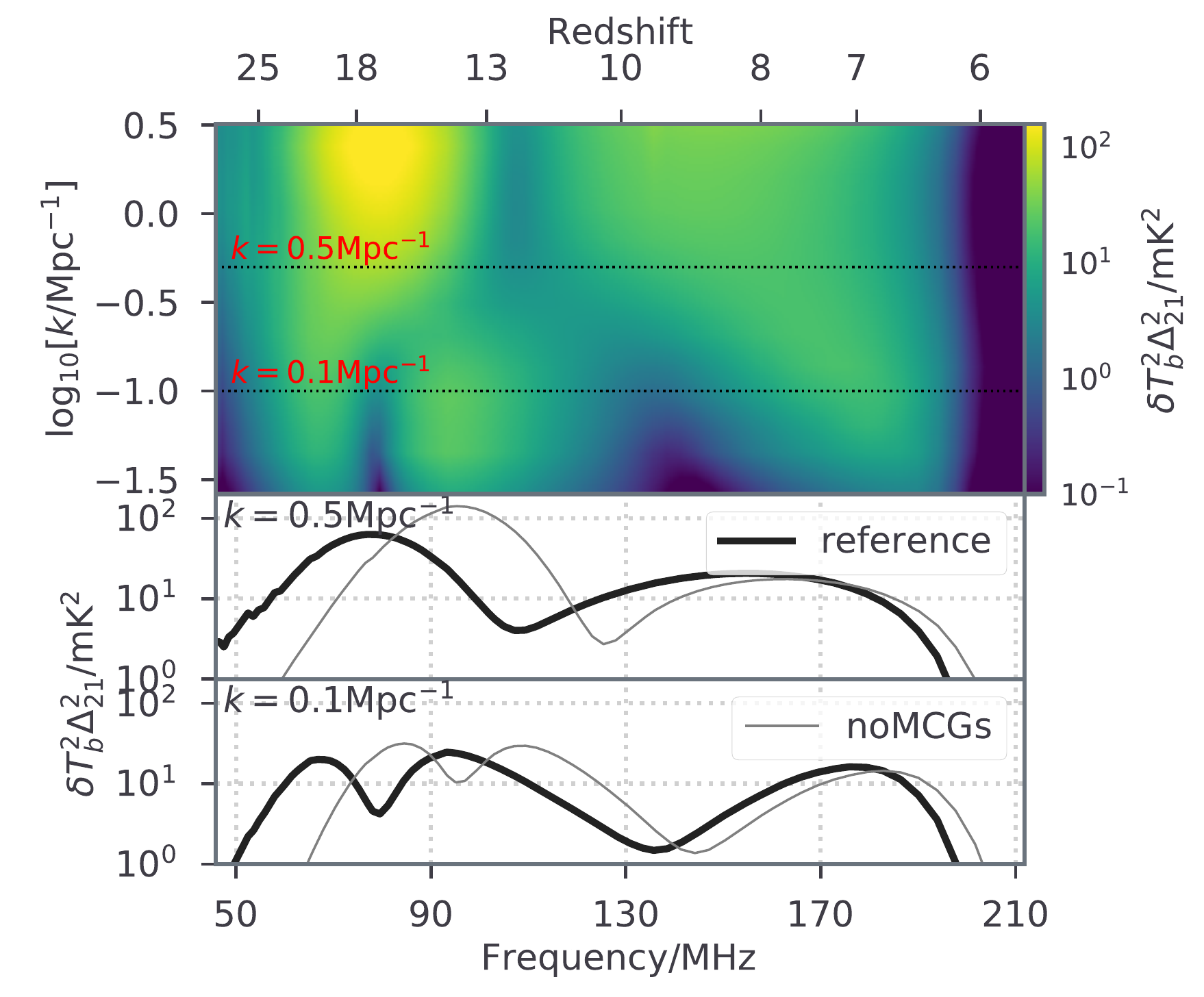}\vspace*{-2mm}
		\end{center}
		\caption{\label{fig:reference_PS}Dimensionless power spectra of 21-cm ($\delta T_{\rm b}^2 \Delta_{21}^2$) for the reference model as a function of wave number and redshift (or observed frequency) on the top panel and for two typical scales in the observable range, $k=0.1$ and 0.5 Mpc$^{-1}$ in the bottom panels. We also present the result from a model without minihaloes (\textit{noMCGs}) for comparison. {\color{black}The kinks at high redshift are numerical due to very rare sources and can be fixed by running the simulation with a higher cadence.}}
	\end{minipage}
\end{figure}

We summarize the relevant model parameters in Table \ref{tab:parameters} together with the values chosen for a reference model. We present this reference model, including slices through various fields in Fig. \ref{fig:reference} and the 21-cm power spectra in Fig. \ref{fig:reference_PS}. Simulations presented in this section share the same initial conditions and are performed within periodic boxes that have a comoving length of 300Mpc and a cell resolution of 1.17Mpc (300Mpc/256). Unless otherwise specified, values are consistent with those in  \citetalias{Park2019MNRAS.484..933P}, for the parameters the two works have in common. We will demonstrate below how current observations can constrain a subset of these parameters in Section \ref{sec:constraint}.

Looking at the light-cones in Fig. \ref{fig:reference}, we see immediately that the structure of the 21-cm signal ({\it rightmost panel}) is governed by various radiation fields, with specific fields dominating at different epochs. The early 21-cm structures ($z\sim20-30$) are imprinted by the {\lya} background ({\it second panel}), which is fairly uniform.  However, regions around the nascent galaxies, which are hosted by large-scale matter overdensities ({\it first panel}), see enhanced fluxes by factors of up to a few. These regions also have a higher LW flux ({\it third panel}), with intensities reaching values large enough for negative feedback on MCGs ({\it seventh panel}) during the {\lya} coupling epoch.  By $z\sim 20$, the LW feedback is significant through the IGM -- the median $J_{\rm LW,eff}^{\rm 21}$ exceeds $10^{-2}$ and the critical mass, $M_{\rm crit}^{\rm diss}$, becomes more than three times the molecular-cooling threshold (see equation \ref{eq:mcrit,diss}), leading to a factor of ${\sim}2$ suppression on the number density of low-mass MCGs.

Shortly thereafter, X-rays from the first galaxies begin to dominate the thermal evolution of the IGM ({\it fourth panel}). By $z\sim18$, $\varepsilon_{X}$ exceeds ${\rm}100 k_{\rm B}{\rm K/Gyr}$ in most parts of the simulation box, overcoming adiabatic cooling of the gas (see equation \ref{eq:T_g}). With $\dot{T}_{\rm g}$ becoming positive, $\delta T_{\rm b}$ reaches its minimum and we see an absorption feature in the 21-cm light-cone, which fades away at $z{\sim}12$ (see also Fig. \ref{fig:reference_PS}). After that, gas becomes hotter than the CMB and the signal is in emission.

\begin{table*}
	\caption{A list of the main relevant parameters in the model with descriptions, values adopted for the reference model, and range for exploration during the MCMC.}
	\vspace{-3mm}
	\begin{threeparttable}
		\label{tab:parameters}
		\begin{tabular}{lcclcc}
			\hline \hline
			\vspace{3mm}
			Parameter & Sec. & Eq. & Description & Ref. & MCMC \\
			\hline
			$\log_{10}f_{*,10}^{\rm atom}$ &\ref{subsec:source model}& \ref{eq:m*_atom}& \multirow{2}{1.9cm}{Stellar-to-halo mass ratio} \multirow{2}{4.5cm}{at 
				$M_{\rm vir}=\begin{array}{l}10^{10}\msol \\ 10^{7}\msol \end{array}$for$\begin{array}{l}\rm ACGs \\ \rm MCGs\end{array}$}& -1.25&-1.25\\
			\vspace{1mm}
			$\log_{10}f_{*,7}^{\rm mol}$ &\ref{subsec:source model_mcg}& \ref{eq:m*_mol}& & -1.75&[-3, 0]\\\hdashline
			\vspace{1mm}
			$\alpha_*$ &\ref{subsec:source model},\ref{subsec:source model_mcg}& \ref{eq:m*_atom},\ref{eq:m*_mol}& Stellar-to-halo mass power-law  index& 0.5&0.5\\\hdashline
			\vspace{1mm}
			$t_*$&\ref{subsec:source model} & \ref{eq:sfr} & Star formation time-scale in units of $H^{-1}(z)$& 0.5&0.5\\\hdashline
			\vspace{1mm}
			$M_{\rm crit}^{\rm SN}$&\ref{subsec:source model},\ref{subsec:source model_mcg}&\ref{eq:m_crit_atom},\ref{eq:m_crit_mol}&Critical halo mass for supernova feedback\tnote{a}&-&-\\\hdashline
			\vspace{1mm}
			$f_{\htwo}^{\rm shield}$&\ref{subsec:source model} & \ref{eq:fshiled} & Self-shielding factor of $\htwo$ for LW dissociation & 0.0&0.0 \\\hdashline
			\vspace{1mm}
			$n_{\gamma}^{\rm atom}$&\ref{subsubsec:UVionizing} & \ref{eq:dotn_ion} & \multirow{2}{4.2cm}{Number of ionizing photons emitted per stellar baryon} \multirow{2}{1.8cm}{for $\begin{array}{l}\rm ACGs \\ \rm MCGs\end{array}$}& $5{\times}10^3$&$5{\times}10^3$\\
			\vspace{1mm}
			$n_{\gamma}^{\rm mol}$&\ref{subsubsec:UVionizing} & \ref{eq:dotn_ion} & & $5{\times}10^4$&$5{\times}10^4$\\\hdashline
			\vspace{1mm}
			$\log_{10}f_{\rm esc,10}^{\rm atom}$ &\multirow{2}{6mm}{\ref{subsubsec:UVionizing}}& \multirow{2}{3mm}{\ref{eq:f_esc}}& \multirow{2}{2.4cm}{Escape fraction of ionizing photons} \multirow{2}{4.5cm}{at 
				$M_{\rm vir}=\begin{array}{l}10^{10}\msol \\ 10^{7}\msol \end{array}$for$\begin{array}{l}\rm ACGs \\ \rm MCGs\end{array}$}&  -1.22&[-3, 0] \\
			\vspace{1mm}
			$\log_{10}f_{\rm esc,7}^{\rm mol}$ &&& & -2.22&[-3, 0]\\\hdashline
			\vspace{1mm}
			$\alpha_{\rm esc}^{\rm atom}$ &\multirow{2}{6mm}{\ref{subsubsec:UVionizing}}& \multirow{2}{3mm}{\ref{eq:f_esc}}& \multirow{2}{5cm}{Escape fraction of ionizing photons to halo mass power-law indices} \multirow{2}{3cm}{for{$\begin{array}{l}\rm ACGs \\ \rm MCGs\end{array}$}}&  0& 0 \\
			\vspace{1mm}
			$\alpha_{\rm esc}^{\rm mol}$ &&& & 0&0\\\hdashline
			\vspace{1mm}
			$\alpha_{\rm UVB}$ & \ref{subsubsec:UVionizing} & \ref{eq:gamma} & Spectral index of the ionizing background & 5&5\\\hdashline
			\vspace{1mm}
			$E_0/{\rm eV}$ & \ref{subsubsec:xraysandlya} & \ref{eq:xspec} & Minimum X-ray energy escaping the galaxies into the IGM & 500&[100, 1500]\\\hdashline
			\vspace{1mm}
			$\alpha_{\rm X}$ & \ref{subsubsec:xraysandlya} & \ref{eq:xspec} & Spectral index of X-ray sources\tnote{b} & 1.0 &1.0\\\hdashline
			\vspace{1mm}
			$\log_{10}L_{\rm X<2keV/\dot{\odot}}^{\rm atom}$ &\multirow{2}{6mm}{\ref{subsubsec:xraysandlya}}& \multirow{2}{3mm}{\ref{eq:xspec}}& \multirow{2}{4.2cm}{Soft-band X-ray luminosity per SFR in units of $\rm erg\ s^{-1} \msol^{-1} yr$} \multirow{2}{2cm}{for$\begin{array}{l}\rm ACGs \\ \rm MCGs\end{array}$}& 40.5&\multirow{2}{8mm}{[38,44]\tnote{c}}  \\
			\vspace{1mm}
			$\log_{10}L_{\rm X<2keV/\dot{\odot}}^{\rm mol}$ &&&&40.5&\\	
			\hline 
			\hline
		\end{tabular}
		\begin{tablenotes}
			\item[a] Although it is a free parameter, for this work we maximize the importance of small galaxies by assuming supernova feedback is subdominant compared to inefficient cooling and photoheating in determining the faint turnover, i.e. $M_{\rm crit}^{\rm SN} {\le} \max\left[M_{\rm crit}^{\rm cool},  M_{\rm crit}^{\rm ion} \right]$;
			\item[b] In this work, we set $\alpha_{\rm X}=1$, motivated by observations of (population-averaged) spectra of high-mass X-ray binaries in local galaxies (e.g. \citealt{Mineo2012,Fragos2013ApJ...764...41F,Pacucci2014}).
			\item[c] We assume that ACGs and MCGs possess similar X-ray luminosities during MCMC, i.e. $L_{\rm X<2keV/\dot{\odot}}^{\rm atom} {=}L_{\rm X<2keV/\dot{\odot}}^{\rm mol}{\equiv} L_{\rm X<2keV/\dot{\odot}}$.
		\end{tablenotes}
	\end{threeparttable}
\end{table*}

The EoR ({\it second to last panel}), as well as the associated photoheating feedback ({\it sixth panel}) and recombinations fields ({\it eight panel}), is driven by short mean free path ionizing photons. Therefore, their evolution is not sensitive to a diffuse, increasing background (as is the case for X-rays and LW photons) but proceeds in a ``percolating fashion'' (e.g. \citealt{furlanetto2016reionization}) with medium-to-large scales being closely tied to the underlying density field (e.g. \citealt{Zahn2011MNRAS.414..727Z,battaglia2013reionization,mcquinn2018observable}).  The EoR history of this model is chosen to agree with current observational constraints, finishing by $z\sim6$ (e.g. \citealt{McGreer2015MNRAS.447..499M}), having a mid-point of around $z\sim7-8$ (\citealt{Planck2016A&A...596A.108P}), and a small tail extending to higher redshifts corresponding to small HII regions around the nascent first galaxies (e.g. \citealt{Mitra2015,Greig2016}).

The 21-cm PS of this model, presented in Fig. \ref{fig:reference_PS}, shows the characteristic triple peak structure of the large-scale power evolution, driven by fluctuations in the {\lya} coupling, X-ray heating, and reionization fields.  On smaller scales, the first two peaks merge due to a larger negative contribution of the cross-terms of the component fields (see discussions in \citealt{Pritchard2007MNRAS.376.1680P, Baek2010A&A...523A...4B, Mesinger2013MNRAS.431..621M}). For reference, we also show in grey the same astrophysical model but with no minihaloes.  We see in general that the astrophysical epochs in this model are delayed, especially the earliest ones, and there is more power on large scales.  We will return to this below.

\begin{figure*}
	\begin{minipage}{\textwidth}
		\begin{center}
			\includegraphics[width=0.98\textwidth]{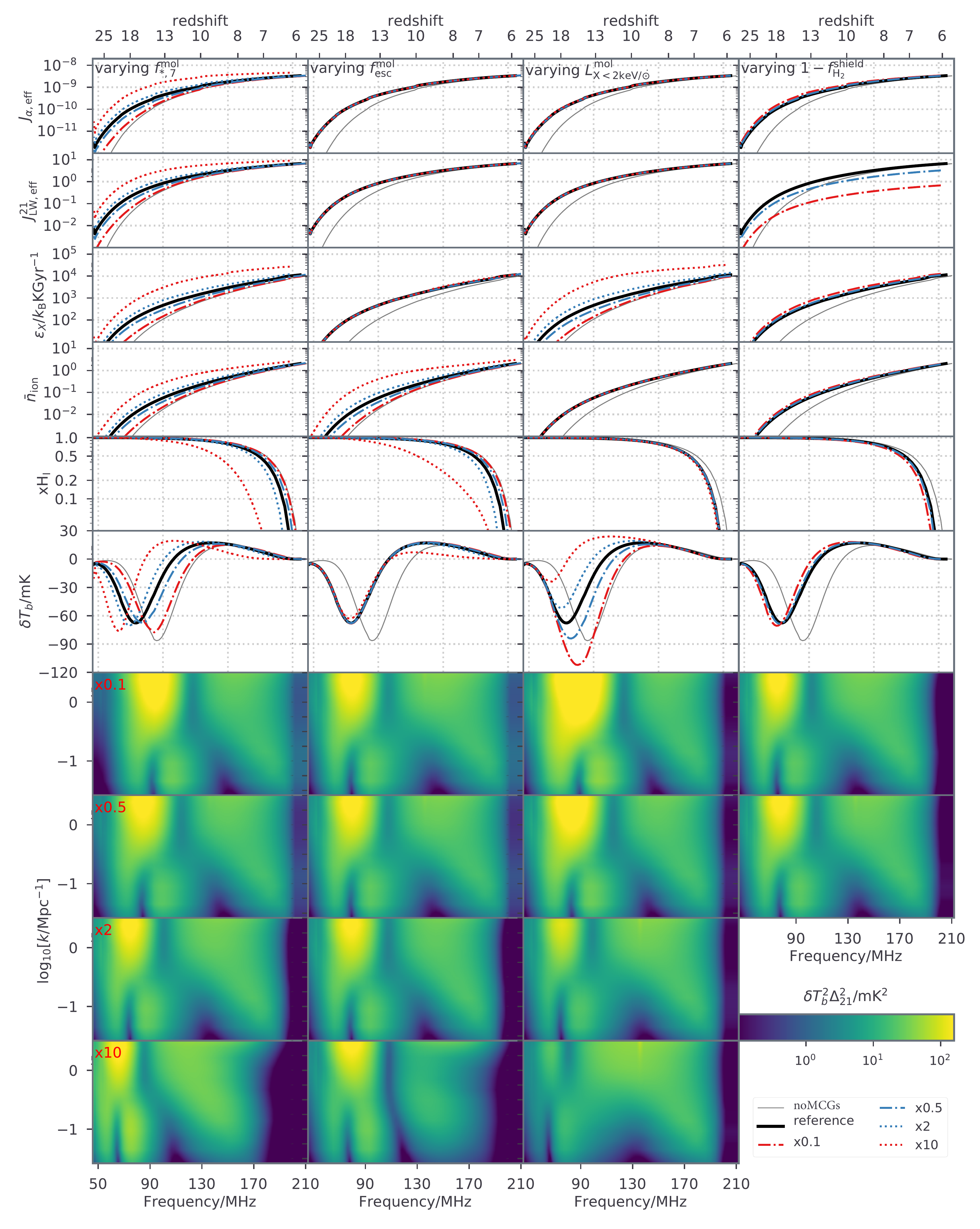}\vspace*{-5mm}
		\end{center}
		\caption{\label{fig:showcases_signal} Varying some of the parameters that describe MCGs by factors of 0.1, 0.5, 2, and 10. From left to right, we show the stellar-to-halo mass ratio at $M_{\rm vir}{=}10^7\msol$ ($f_{*,7}^{\rm mol}$), ionizing escape fraction ($f_{\rm esc}^{\rm mol}$), X-ray luminosity per SFR ($L_{\rm X<2keV/\dot{\odot}}^{\rm mol}$), and self-shielding factor of {$\htwo$} ($f_{\htwo}^{\rm shield}$). {\color{black}Note that varying $1{-}f_{\htwo}^{\rm shield}$ by a factor of 0.1 and 0.5 results in $f_{\htwo}^{\rm shield}{=}0.9$ and 0.5}. The first four rows present the evolution of radiation fields including {\lya} ($J_{\alpha, {\rm eff}}$), LW ($J^{\rm 21}_{\rm LW,eff}$), X-ray heating ($\varepsilon_{X}$) and the cumulative number of ionizing per baryon ($\bar{n}_{\rm ion}$). The next two rows correspond to the neutral hydrogen fraction ($x_{\hone}$) and average 21-cm brightness temperature ($\delta T_{\rm b}$). We show the corresponding dimensionless power spectra ($\delta T_{\rm b}^2 \Delta_{21}^2$) of 21-cm as a function of wave number and redshift on the last four rows. We also present the result from a model  without minihaloes ({\it noMCGs}) for comparison. }
	\end{minipage}
\end{figure*}

\subsection{Parameter dependence}

Here we vary some of the astrophysical parameters characterizing MCGs, illustrating the resulting impact on cosmic fields.  In Fig. \ref{fig:showcases_signal}, we show the redshift evolution of the median values of various fields in the top rows, together with the 21-cm PS in the bottom rows.  Along the columns, we vary the normalization of the stellar-to-halo mass relation (i.e. ratio at $M_{\rm vir}{=}10^7\msol$; $f_{*,7}^{\rm mol}$), the ionizing escape fraction ($f_{\rm esc}^{\rm mol}$), the soft-band X-ray luminosity per SFR ($L_{\rm X<2keV/\dot{\odot}}^{\rm mol}$) and self-shielding factor of {$\htwo$} ($f_{\htwo}^{\rm shield}$).  We only vary one parameter at a time, fixing the remaining parameters to those of the reference model. The general trends are the following:
\begin{enumerate}
	\item varying $f_{*,7}^{\rm mol}$ leads to different production rates of photons in all wavelengths of interest. An increasing stellar mass in MCGs shifts all astrophysical epochs to earlier times.  Understandably, the earlier epochs of {\lya} pumping and X-ray heating are most {\color{black}affected}, as MCGs have a larger relative contribution at higher redshifts. However increasing the efficiency to 10 times our fiducial one (or analogously increasing the ionizing escape fraction) shifts reionization to earlier times.  In this case, MCGs can contribute more ionizing photons than ACGs well into the EoR ($z\gsim6$; comparing grey and red dotted curves in the fourth row), and the midpoint of the EoR shifts to $z\sim9$;
  
	\item varying $f_{\rm esc}^{\rm mol}$ around our fiducial model only has a minor impact on the timing of the EoR. Because the overall emission of ionizing photons depends on the product of the escape fraction and the SFR, increasing the escape fraction by a factor of 10 results in a shift of the EoR to earlier times, as seen in the previous column.  Note that the stellar mass and the escape fraction do not have a completely degenerate impact on the EoR timing as the star formation time-scale evolves with redshift, and radiative feedback can regulate star formation;

	\item varying $L_{\rm X<2keV/\dot{\odot}}^{\rm mol}$ impacts almost exclusively the Epoch of Heating (EoH), as X-rays are inefficient at reionizing the Universe.  Increasing the X-ray luminosity shifts the EoH to earlier times.  As a result, the EoR and {\lya} pumping epochs increasingly overlap, which dramatically reduces the maximum contrast between the gas and CMB temperatures, and the corresponding minima in the global signal.  Moreover, the resulting 21-cm power on small and medium scales is also reduced due to the increased negative contribution of the cross-correlation between the temperature  and {\lya} coupling fields;\footnote{In other words, during {\lya} pumping, the regions close to galaxies have the strongest coupling, with their spin temperatures approaching the gas temperature ($T_s \sim T_g \ll T_{\rm cmb}$) while most of the IGM has a spin temperature close to that of the CMB ($T_s \sim T_{\rm cmb} \gg T_g$).  Thus regions close to galaxies appear as cold spots in the 21-cm signal during this early stage when the IGM is still colder than the CMB.  However, if X-ray heating is more efficient, the gas surrounding the first galaxies can heat up before coupling is completed.  In such a case of strong overlap of the EoH and epoch of {\lya} coupling, regions close to galaxies can be heated and coupled ($T_s = T_g \sim T_{\rm cmb}$), while those regions distant from galaxies are still cold but not coupled ($T_s \sim T_{\rm cmb} \gg T_g$).  In this case most of the IGM can have spin temperatures that are closer to the CMB temperature, reducing the mean 21-cm signal and spatial fluctuations.}

	\item varying $f_{\htwo}^{\rm shield}$ changes how sensitive MCGs respond to negative feedback from the LWB (e.g. \citealt{Schauer2015MNRAS.454.2441S}). As our reference model assumes no self-shielding, $f_{\htwo}^{\rm shield} = 0$, here we only increase this parameter to 0.5 and 0.9. A larger $f_{\htwo}^{\rm shield}$ decreases the effective LW radiation penetrating the ISM of the galaxies ({\it second row}), decreasing the impact of LW feedback. With a correspondingly higher contribution of MCGs when self-shielding is increased, astrophysical epochs are shifted earlier; however, the effect is extremely small, indicating that negative LW feedback in our model is not very important.
\end{enumerate}

We also present a model (\textit{noMCGs}) where contribution from minihaloes is excluded. Comparing with \textit{noMCGs}, we see that, depending on the values used for the aforementioned parameters, MCGs can be the dominant source of radiation in the early universe, governing the global evolution of 21-cm signal, and altering its morphology. Therefore, 21-cm observables can be a powerful tool to probe the properties of first galaxies. In the next section, we will quantify how current high-redshift observations can jointly constrain the properties of MCGs and ACGs within a Bayesian analysis framework.

\section{Inferring the astrophysics of minihaloes}\label{sec:constraint}

The previous section illustrates how varying galaxy properties can impact the 21-cm signal.  However, our model has many free parameters which characterize both ACGs and MCGs. Can these parameters be constrained by current and upcoming observations?  
In a companion paper, we will quantify the parameter constraints and degeneracies available with future 21-cm interferometric observations. Here we focus on current observations of the EoR and CD, seeing if these can already be used to inform our model and infer the astrophysics of minihaloes.  These observations \footnote{We assume the corresponding uncertainties to be Gaussian or one-sided Gaussian (for upper limits).} include
\begin{enumerate}
	\item the galaxy UV LF at $z{=}6{-}10$ from \citet{Bouwens2015a,Bouwens2016} and \citet{Oesch2018ApJ...855..105O};
	\item the upper limit on the neutral hydrogen fraction at $z\sim5.9$, $x_{\hone}<0.06{+}0.05(1\sigma)$, measured using the dark fraction of QSO spectra \citep{McGreer2015MNRAS.447..499M};
	\item the Thomson scattering optical depth of CMB photons reported by \citet{Planck2016A&A...596A.108P}, $\tau_e=0.058{\pm}0.012(1\sigma)$; and
	\item the timing\footnote{If the EDGES signal is indeed cosmological, its amplitude could only be explained with non-standard models (e.g. \citealt{Ewall-Wice2018ApJ...868...63E,Fialkov2018PhRvL.121a1101F,Munoz2018Natur.557..684M,Mebane2019arXiv191010171M}).  We do not go into the physical sources of the unexpectedly deep absorption signal in this work.  Nevertheless, current explanations still rely on X-rays and soft UV radiation from galaxies to govern its {\it timing}.} of the 21-cm global absorption feature reported by EDGES, which has a minimum at a frequency of 78$\pm1(1\sigma)$MHz \citep{Bowman2018}.
\end{enumerate}

To quantify parameter constraints implied by these observations, we use the MCMC module, 21CMMC \citep{Greig2015MNRAS.449.4246G,Greig2017MNRAS.472.2651G,Greig2018MNRAS.477.3217G}, which forward-models 21-cm light-cones using the EMCEE sampler \citep{Goodman2010CAMCS...5...65G,Foreman2013PASP..125..306F}. Unfortunately, varying all of the model parameters listed in Table 1 is computationally challenging, and would require high-performance computing resources. We defer a larger parameter space exploration to future work.

For this introductory work, we limit our parameter space. Specifically, we fix the stellar-to-halo mass relation of ACGs to the recovered median values in \citetalias{Park2019MNRAS.484..933P}, as current LFs already provide reasonable constraints on these parameters.
These include: (i)) the stellar-to-halo mass ratio at $M_{\rm}=10^{10}{\rm M}_\odot$ for ACGs, $\log_{10} f_{*,10}^{\rm atom}{=}{-}1.25$; (ii) the power-law index of the ACG stellar-to-halo mass relation, $\alpha_{*}{=}0.5$; and (iii) the star formation time-scale, $t_* {=} 0.5$. Fixing these values ensures that the modelled galaxy UV LFs are in agreement with high-redshift observations at the bright end (see also Fig. \ref{fig:example}).

\begin{figure*}
	\begin{minipage}{\textwidth}
		\begin{center}
			\includegraphics[width=\textwidth,valign=t]{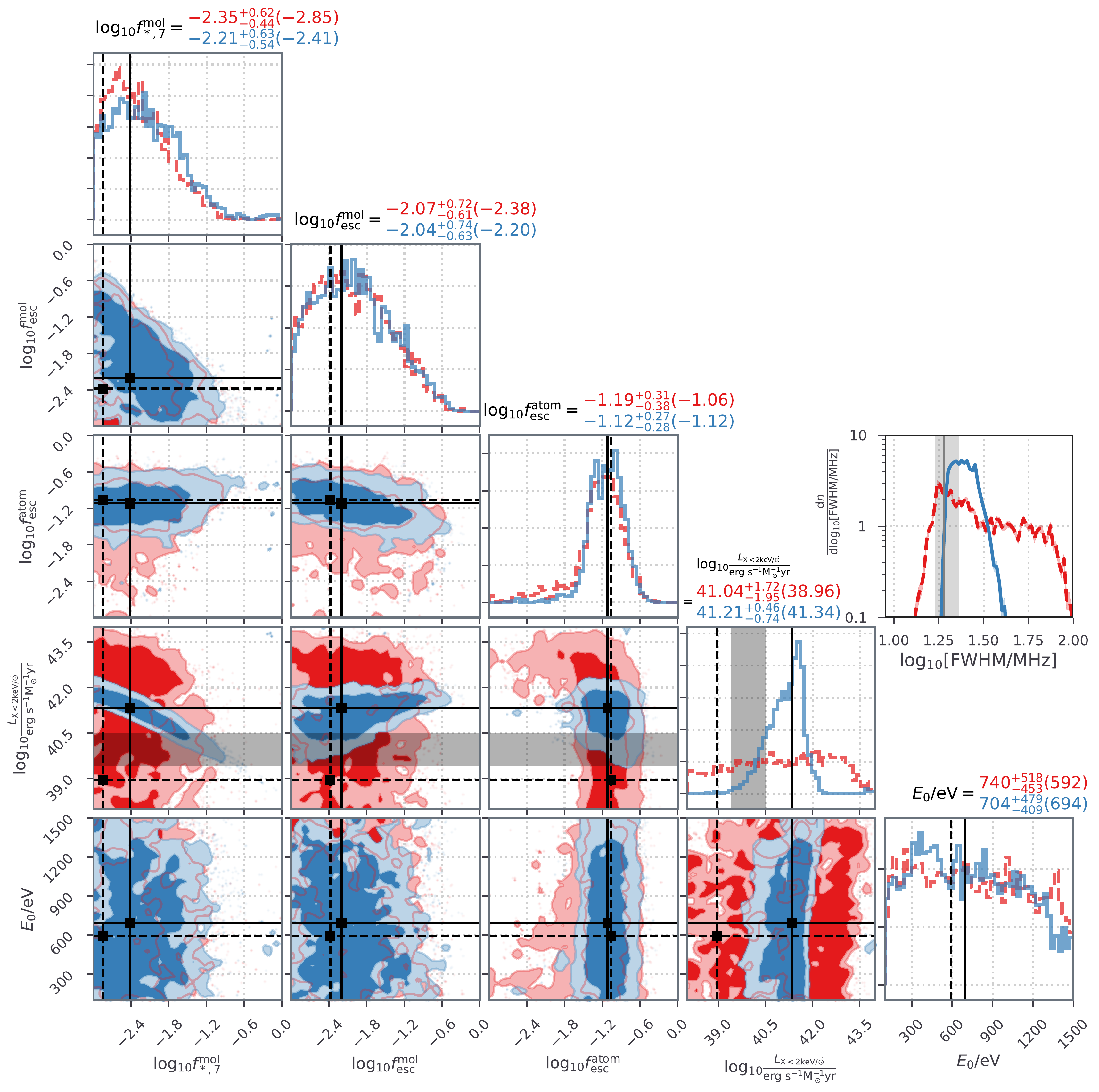}\hspace*{-0.52\textwidth}
			\includegraphics[width=0.52\textwidth,valign=t]{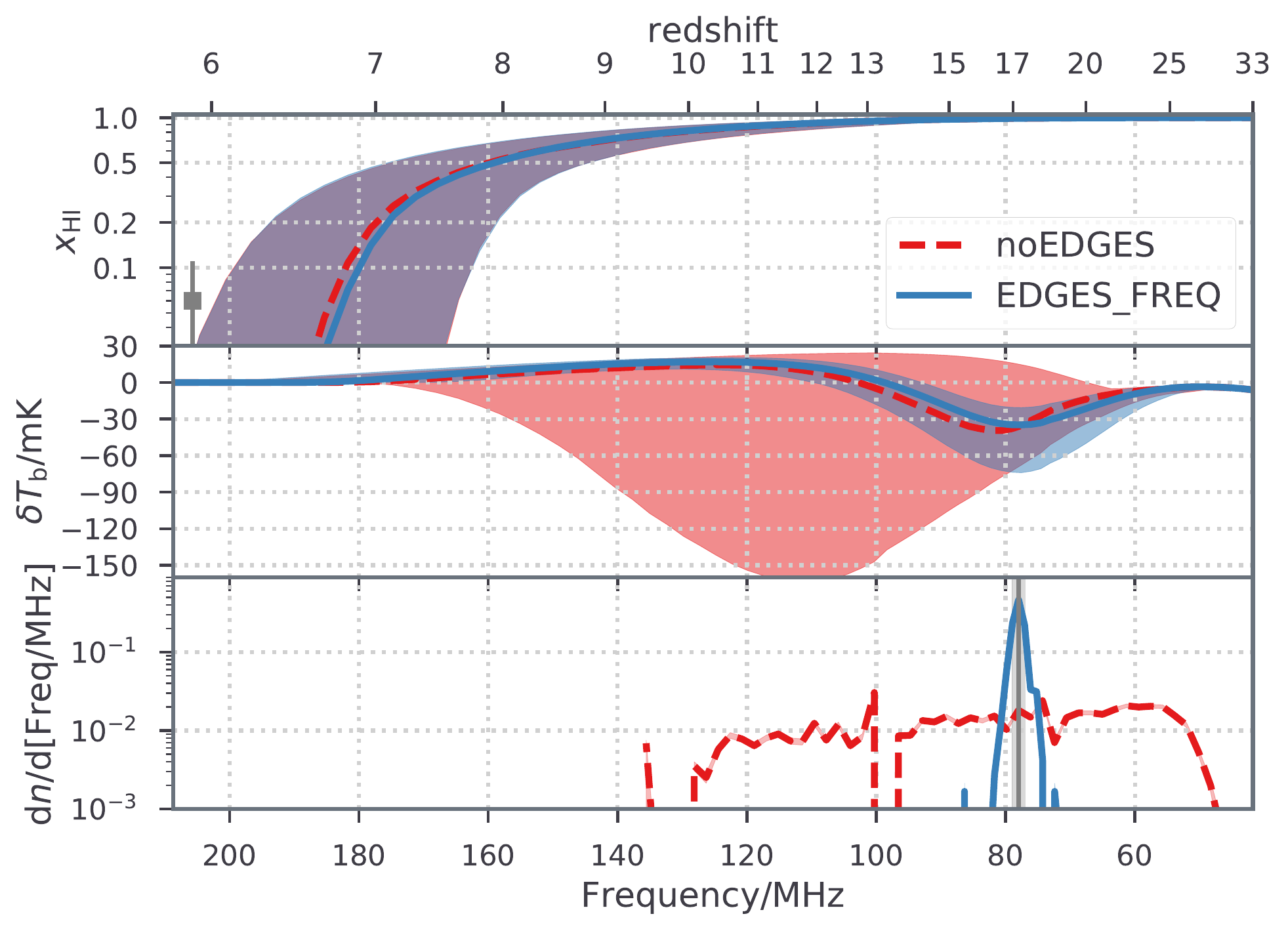}
		\end{center}
	\end{minipage}
	\caption{\label{fig:MCMC}Marginalized posterior distributions of the model parameters with different observational constraints (\textit{noEDGES} in red or using dashed lines, \textit{EDGES\_FREQ} in blue or using solid lines). While both results use the observed galaxy LFs at $z\sim6{-}10$ \citep{Bouwens2015a,Bouwens2017ApJ...843..129B,Oesch2018ApJ...855..105O}, upper limits on the neutral fraction at $z{\sim}5.9$ from QSO spectra \citep{McGreer2015MNRAS.447..499M} and Thomson scattering optical depth of the CMB \citep{Planck2016A&A...596A.108P}, \textit{EDGES\_FREQ} includes an additional constraint from the timing of the absorption in the sky-averaged 21-cm spectrum \citep{Bowman2018}. The {\color{black} 68th and 95th percentiles} are indicated using dark and light shaded regions in the 2D distributions, respectively. The medians with [14,86] percentiles are presented on the top of the 1D PDF for each parameter. The models located at the maximum likelihood are indicated by black solid and dashed lines on the PDFs for \textit{EDGES\_FREQ} and \textit{noEDGES}, respectively, with their parameter values shown in brackets. We observe degeneracies between $\log_{10}f_{*,7}^{\rm mol}$ and $\log_{10}f_{\rm esc}^{\rm mol}$ with their sum being ${-4.23_{-0.74}^{+0.58}}$ (median with [14,86] percentiles; \textit{noEDGES}) and ${-4.07_{-0.73}^{+0.50}}$ (\textit{EDGES\_FREQ}) as well as strong degeneracies between $\log_{10}f_{*,7}^{\rm mol}$ and $\log_{10}\left(L_{\rm X<2keV/\dot{\odot}}/{\rm erg\ s^{-1}\ M_{\odot}^{-1}\ yr}\right)$ in \textit{EDGES\_FREQ} with their sum being $38.92_{-0.15}^{+0.24}$. In marginalized posteriors of $\log_{10}\left(L_{\rm X<2keV/\dot{\odot}}/{\rm erg\ s^{-1}\ M_{\odot}^{-1}\ yr}\right)$, the grey shaded region corresponds to the mean from HMXBs in local star-forming galaxies ({\it lower bound}; e.g. \citealt{Mineo2012}) and a factor of 10 enhancement theoretically expected in a metal-poor environment ({\it upper bound}; e.g. \citealt{Fragos2013ApJ...764...41F}); note that we conservatively do not use this as a prior in our MCMC. On the upper right corner, the median and [14,86] percentiles of the neutral hydrogen ($x_{\hone}$) and brightness temperature ($\delta T_{\rm b}$) cosmic evolution are shown for the models presented in the posterior distributions. Corresponding PDFs of the absorption frequency in $\delta T_{\rm b}\left({\rm frequency}\right)$ and its FWHM are presented with observational constraints ($1\sigma$; \citealt{Bowman2018}) shown in grey.}
\end{figure*}

Additionally, we consider constant escape fractions for each population (i.e. ${\alpha_{\rm esc}^{{\rm atom}({\rm mol})}}=0$), ignore self-shielding of $\htwo$ (i.e. $f_{\htwo}^{\rm shield}=0$), and further assume ACGs and MCGs possess a similar X-ray luminosity per SFR\footnote{Although the X-ray luminosity of HMXBs scales with decreasing metallicity (\citealt{Mapelli2010MNRAS.408..234M,Douna2015A&A...579A..44D,Brorby2016MNRAS.457.4081B}), theoretically this trend is expected to saturate for metallicities below roughly 10 per cent solar (e.g. \citealt{Fragos2013ApJ...764...41F}).  Thus, assuming similar X-ray luminosity to SFRs for ACGs and MCGs could be reasonable if the level of metal enrichment in early ACGs is fairly modest.  In any case, our results can be treated as a lower limit on the contribution of MCGs to the X-ray background.} (i.e. $L_{\rm X<2keV/\dot{\odot}}^{\rm atom} = L_{\rm X<2keV/\dot{\odot}}^{\rm mol}\equiv L_{\rm X<2keV/\dot{\odot}}$).

We thus explore the following parameters with flat priors in linear or logarithmic scale:
\begin{enumerate}
	\item the normalization of the MCG stellar-to-halo mass ratio, $\log_{10} f_{*,7}^{\rm mol}\in\left[-3,0\right]$;
	\item the escape fraction of ionizing photons for ACGs, $\log_{10} f_{\rm esc}^{\rm atom}\in\left[-3,0\right]$;
	\item the escape fraction of ionizing photons for MCGs, $\log_{10} f_{\rm esc}^{\rm mol}\in\left[-3,0\right]$;
	\item the minimum energy for X-rays to reach the IGM, $E_0\in\left[100,1500\right]{\rm eV}$; and
	\item the soft-band X-ray luminosity per SFR, $\log_{10}\left[L_{\rm X<2keV/\dot{\odot}}/{\rm erg\ s^{-1}\msol^{-1} yr}\right]\in\left[38, 44\right]$.
\end{enumerate}
For the sake of computing efficiency, we have chosen a slightly smaller box with a comoving length of 250Mpc and a cell resolution of 1.95Mpc (250Mpc/128) when performing the MCMC.
 
Fig. \ref{fig:MCMC} shows the marginalized posterior distributions together with the corresponding marginalized [14, 86] percentiles of the average EoR and 21-cm redshift evolutions.  We also identify the timing when $\delta T_{\rm b}$ reaches its minimum as well as the full width at half-maximum (FWHM) of $\delta T_{\rm b}-$frequency, and show their PDFs in the right-hand subpanels. The red curves and shaded areas correspond to constraints using all of the above observations, {\it except} EDGES (\textit{noEDGES}).  Even without EDGES, we see a strong degeneracy between the allowed SFR and the ionizing escape fractions in MCGs -- high values of either $f_{*,7}^{\rm mol}$ or $f_{\rm esc}^{\rm mol}$ are excluded, as they would reionize the Universe too early to be consistent with {\it Planck} observations (see also, e.g. \citealt{Visbal2015MNRAS.453.4456V}). 
On the other hand, an escape fraction of ionizing photons in ACGs of $f_{\rm esc}^{\rm atom}{\sim}3 {-}15$\% is required to ensure a sufficiently ionized universe at $z{\sim}6$. As expected, without any information of 21cm, the X-ray properties cannot be constrained by any of these measurements.

We then add in the constraints from EDGES timing (\textit{EDGES\_FREQ}). The corresponding marginalized PDFs are shown with blue curves and shaded regions. Most constraints tighten only slightly when including the timing of the EDGES signal.
In particular, the aforementioned degeneracy between the stellar-halo mass ratio and ionizing escape fraction is mostly unchanged.  We fit this degeneracy in both cases to obtain the following relations (median with [14, 86] percentiles):
\begin{equation}\label{eq:f*fesc}
\log_{10}\left({f_{*,7}^{\rm mol}f_{\rm esc}^{\rm mol}}\right) = 
\begin{cases}
-4.23_{-0.74}^{+0.58} ~ ({\it noEDGES})\\
-4.07_{-0.73}^{+0.50} ~ ({\it EDGES\_FREQ}).
\end{cases}
\end{equation}
However, the most striking change is in the $f_{*,7}^{\rm mol}$ -- $L_{\rm X<2keV/\dot{\odot}}$ plane. We see that a strong degeneracy emerges between these two parameters
\begin{equation}\label{eq:f*lx}
 \log_{10}\left(f_{*,7}^{\rm mol}\frac{L_{\rm X<2keV/\dot{\odot}}}{ {\rm erg\ s^{-1}\ M_{\odot}^{-1}\ yr}} \right) = 38.92_{-0.15}^{+0.24} ~ ({\it EDGES\_FREQ}).
\end{equation}
If the EDGES signal at $78\pm1$MHz is cosmological, soft UV and X-ray photons from galaxies are needed at $z\sim$17--20 to source the {\lya} coupling and subsequent X-ray heating transitions, regardless of the physical explanation of the depth of the signal \citep{Madau2018,Mirocha2019MNRAS.483.1980M}. However, the stellar-to-halo mass relation implied by observations of high-redshift UV LFs is insufficient to heat the IGM at such high redshifts \citep{Mirocha2016,Mirocha2019MNRAS.483.1980M,Park2020MNRAS.491.3891P}. This is in contrast with early estimates of X-ray heating, based on assumptions of a constant stellar-to-halo mass ratio (e.g. \citealt{Mesinger2016,Fialkov2018PhRvL.121a1101F}). As a result, the cosmological explanation of the EDGES signal requires MCGs to set the timing of the signal.\footnote{One could get around this claim if the ACGs were allowed to have higher values for the X-ray luminosity-to-SFR relation, i.e. $L_{\rm X<2keV/\dot{\odot}}^{\rm atom} > L_{\rm X<2keV/\dot{\odot}}^{\rm mol}$.  However, this is contrary to the expected trend, since $L_{\rm X<2keV/\dot{\odot}}$ for HMXBs should {\it increase} with {\it decreasing} metallicity (e.g. \citealt{Mapelli2010MNRAS.408..234M,Douna2015A&A...579A..44D,Brorby2016MNRAS.457.4081B}).}

\begin{figure*}
	\begin{minipage}{\textwidth}
		\begin{center}
			\includegraphics[width=0.65\textwidth]{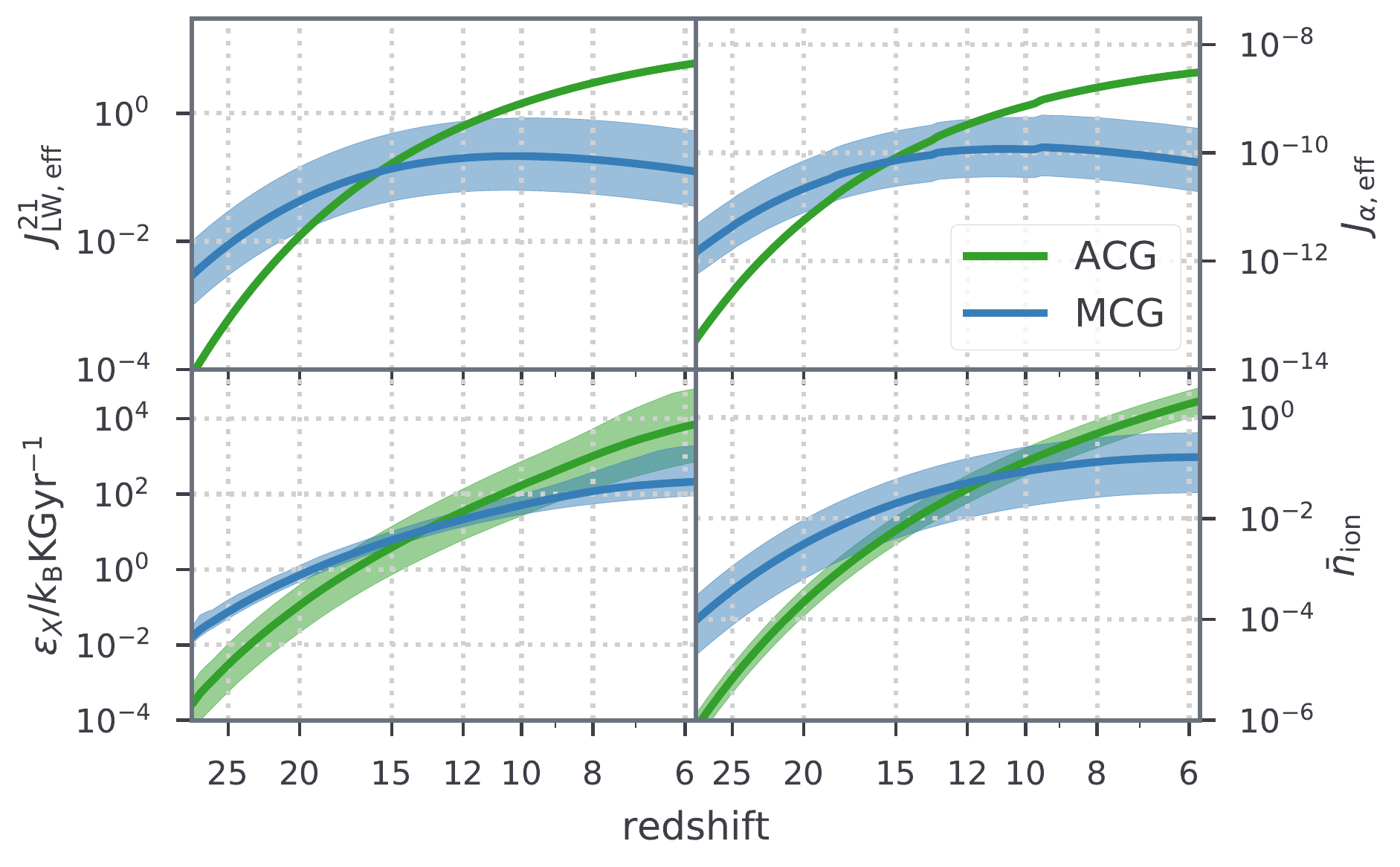}\vspace*{-2mm}
		\end{center}
		\caption{\label{fig:MCMC_photonbudget} Evolution of radiation background including LW ($J^{\rm 21}_{\rm LW,eff}$), {\lya} ($J_{\alpha, {\rm eff}}$), UV ionizing ($\bar{n}_{\rm ion}$) and X-ray heating ($\varepsilon_{X}$) from the \textit{EDGES\_FREQ} posterior distribution as shown in Fig. \ref{fig:MCMC}. The contribution from ACGs and MCGs is indicated in green and blue, respectively, with line and shaded region representing the median and [16, 84] percentiles. Note that the narrow distribution for ACGs is caused by fixing $f_{*,10}^{\rm atom}$ and $\alpha_{*}$. MCGs dominate the LW, {\lya}, UV ionizing and X-ray radiation at $z{\gtrsim}15$, ${\gtrsim}15$, ${\gtrsim}11$, and ${\gtrsim}13$, respectively.}
	\end{minipage}
\end{figure*}

We quantify this claim further in Fig. \ref{fig:MCMC_photonbudget}, which shows the relative contribution of MCGs and ACGs in the LWB, {\lya} background, cumulative ionizing photon number, and X-ray heating rates, corresponding to the \textit{EDGES\_FREQ} posterior\footnote{The contribution of ACGs to $J^{\rm 21}_{\rm LW,eff}$ and $J_{\alpha,{\rm eff}}$ does not have a spread for our posterior, since we are fixing $f_{*,10}^{\rm atom}$ and $\alpha_*$ (motivated by the comparably tight, $\lsim 0.3$ dex constraints on these parameters from current LF observations; \citetalias{Park2019MNRAS.484..933P}), and since we are assuming that SN feedback is subdominant in setting the turnover mass (which maximizes the abundances of ACGs). Varying the X-ray luminosity and escape fraction, however, does result in a spread in the X-ray heating and cumulative number of ionizing photos per baryon for ACGs.}. We see that MCGs dominate the LW, {\lya}, UV ionizing and X-ray radiation at $z{\gtrsim}15$, ${\gtrsim}15$, ${\gtrsim}11$ and ${\gtrsim}13$, respectively, showing that they are the dominant population during the cosmological interpretation of the EDGES signal.

\begin{figure*}
	\begin{minipage}{\textwidth}
		\begin{center}
			\includegraphics[width=0.6\textwidth]{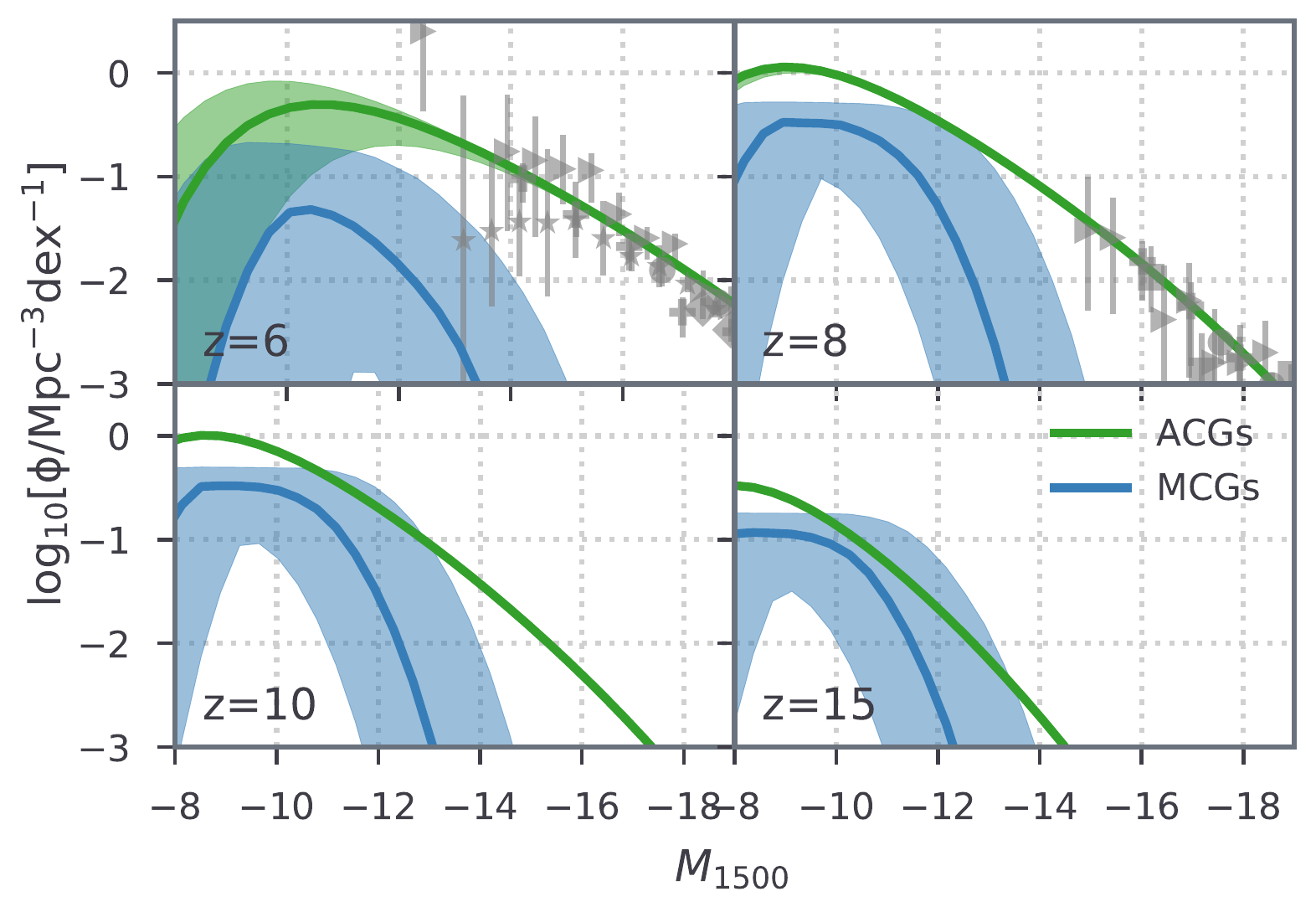}\vspace*{-2mm}
		\end{center}
		\caption{\label{fig:MCMC_LF} UV luminosity functions of MCGs and ACGs from the \textit{EDGES\_FREQ} posterior distribution with lines and shaded regions representing the median and [16,84] percentiles. We note that the scatter in the ACG LFs is underestimated, since we are assuming that SN feedback is subdominant in setting the faint end turnover, and we hold the stellar mass-to-halo mass relation fixed to that recovered in P19.  Observational data at the bright end {\color{black}(diamonds: \citealt{Finkelstein2015ApJ...810...71F}; circles: \citealt{Bouwens2015a}; triangles: \citealt{Livermore2017ApJ...835..113L}; stars: \citealt{Atek2018MNRAS.479.5184A}; pentagons: \citealt{Bhatawdekar2019MNRAS.tmp..843B}; squares: \citealt{Ishigaki2018ApJ...854...73I})} are shown in grey for comparison.}
	\end{minipage}
\end{figure*}

Nevertheless, we note that the MCGs required for explaining EDGES timing are still ``unusual''.  The degeneracies quantified in equations (\ref{eq:f*fesc}) and (\ref{eq:f*lx}) mean that MCGs must have been inefficient at producing ionizing photons but efficient in emitting X-ray photons.  For example, if the ionizing escape fraction of MCGs was above 10 percent (e.g. \citealt{Xu2016ApJ...833...84X}), then in order to match both {\it Planck} and EDGES observations, the star formation (X-ray production) efficiencies of MCGs must have been a factor of $\gsim$ 10 smaller ($\gsim$100-1000 greater) than those of ACGs.  In Fig. \ref{fig:MCMC}, we indicate the expected range for $L_{\rm X<2keV/\dot{\odot}}$ if HMXBs were similar to those in local galaxies \citep{Mineo2012} extrapolated to low-metallicity environments \citep{Fragos2013ApJ...764...41F}.  
We see qualitatively that most MCGs require higher X-ray efficiencies than are theoretically expected even extrapolating to metal-free environments.

Finally, in Fig. \ref{fig:MCMC_LF}, we show the marginalized UV LFs of ACGs\footnote{We see that, by fixing $f_{*,10}^{\rm atom}$, ACG UV LFs only show large uncertainties at low redshifts and faint galaxies, when photoheating feedback from reionization becomes significant. As stated previously, this is due to our assumption of $M_{\rm crit}^{\rm SN} {\le} \max\left[M_{\rm crit}^{\rm cool},  M_{\rm crit}^{\rm ion} \right]$, so as to maximize the star formation in faint galaxies, which are the focus of this work.  Thus, we caution that the scatter in the ACG LFs is underestimated.} and MCGs corresponding to the {\it EDGES\_FREQ} posterior ({\it blue shaded regions}; the {\it noEDGES} LFs are comparable due to the similar distributions of $f_{*,7}^{\rm mol}$ and $f_{\rm esc}$ between these two results). Comparing between LFs of MCGs and ACGs, we recover the result from Fig. \ref{fig:example} and find that MCGs likely only dominate at very high redshifts and faint magnitudes. 

\section{Conclusions}\label{sec:conclusion}
In this work, we include the first, molecularly cooled galaxies that are expected to start the CD in the public {\cmfast} simulation code.  
We consider atomic- (ACGs) and molecular-cooled galaxies (MCGs) as two different populations that source the underlying radiation backgrounds and drive the structure of 21-cm brightness temperature. We allow the stellar mass and SEDs of the two galaxy populations to scale differently with the host halo mass. We track inhomogeneous recombinations and self-consistently follow the relevant radiative feedback mechanisms, including inhomogeneous LW feedback and photoheating feedback on star formation.

We demonstrate how 21-cm observables vary with MCG properties. These include their star formation efficiencies, UV ionizing escape fractions, X-ray luminosities and ${\htwo}$ self-shielding factors against LWB. We then use the Bayesian analysis tool, 21CMMC, to quantify what current observations imply for the MCG population.  We use constraints from: (i) current galaxy luminosity functions at $z{\sim}6{-}10$ \citep{Bouwens2015a,Bouwens2016,Oesch2018ApJ...855..105O}; (ii) the dark fraction upper limit on the neutral hydrogen fraction at $z{\sim}5.9$ \citep{McGreer2015MNRAS.447..499M}; and (iii) the Thomson scattering optical depth of the CMB \citep{Planck2016A&A...596A.108P}. We find that the optical depth already rules out models with a high stellar-to-halo mass ratio and high escape fractions in MCGs \citep{Visbal2015MNRAS.453.4456V}.

We also consider the timing of the first claimed detection of the sky-average 21-cm brightness temperature, from EDGES, as an ancillary data set. We find that MCGs are required to produce a global absorption signal around 78MHz. Moreover, the resulting strong degeneracy between the SFRs and X-ray luminosities of MCGs implies that they would have unexpected properties.  For example, if the ionizing escape fraction of MCGs was above 10 per cent (e.g. \citealt{Xu2016ApJ...833...84X}), then in order to match both {\it Planck} and EDGES observations, the star formation (X-ray production) efficiencies of MCGs must have been a factor of $\gsim$ 10 smaller ($\gsim$100-1000 greater) than those of ACGs. We conclude that the high-redshift 21-cm signal can be a powerful probe of the properties of the first galaxies, which are too faint to be studied using direct observations even with {\it JWST}.

\section*{Acknowledgements}
{\color{black}The authors thank Zoltan Haiman and the anonymous referee for their comprehensive review and positive comments.}
This work was supported by the European Research Council (ERC) under the European Union’s 
Horizon 2020 research and innovation programme (AIDA -- \#638809). Parts of this research 
were supported by the Australian Research Council Centre of Excellence for All Sky 
Astrophysics in 3 Dimensions (ASTRO 3D), through project \#CE170100013 as well as the NSF grant AST-1813694.

\bibliographystyle{\dir mn2e}
\bibliography{reference}

\begin{thebibliography}{170}
\expandafter\ifx\csname natexlab\endcsname\relax\def\natexlab#1{#1}\fi

\bibitem[{{Ahn} {et~al}\mbox{.}(2012){Ahn}, {Iliev}, {Shapiro}, {Mellema},
  {Koda}, \& {Mao}}]{Ahn2012ApJ...756L..16A}
{Ahn} K., {Iliev} I.~T., {Shapiro} P.~R., {Mellema} G., {Koda} J., {Mao} Y.,
  2012, \apjl, 756, L16

\bibitem[{{Ahn} {et~al}\mbox{.}(2009){Ahn}, {Shapiro}, {Iliev}, {Mellema}, \&
  {Pen}}]{Ahn2009ApJ...695.1430A}
{Ahn} K., {Shapiro} P.~R., {Iliev} I.~T., {Mellema} G., {Pen} U.-L., 2009,
  \apj, 695, 1430

\bibitem[{{Atek} {et~al}\mbox{.}(2018){Atek}, {Richard}, {Kneib}, \&
  {Schaerer}}]{Atek2018MNRAS.479.5184A}
{Atek} H., {Richard} J., {Kneib} J.-P., {Schaerer} D., 2018, \mnras, 479, 5184

\bibitem[{{Baek} {et~al}\mbox{.}(2010){Baek}, {Semelin}, {Di Matteo}, {Revaz},
  \& {Combes}}]{Baek2010A&A...523A...4B}
{Baek} S., {Semelin} B., {Di Matteo} P., {Revaz} Y., {Combes} F., 2010, \aap,
  523, A4

\bibitem[{Barkana \& Loeb(2001)}]{Barkana2000}
Barkana R., Loeb A., 2001, Physics Reports, 349, 125

\bibitem[{{Barkana} \& {Loeb}(2005)}]{Barkana2005ApJ...626....1B}
{Barkana} R., {Loeb} A., 2005, \apj, 626, 1

\bibitem[{Battaglia {et~al}\mbox{.}(2013)Battaglia, Trac, Cen, \&
  Loeb}]{battaglia2013reionization}
Battaglia N., Trac H., Cen R., Loeb A., 2013, \apj, 776, 81

\bibitem[{{Beardsley} {et~al}\mbox{.}(2016){Beardsley}, {Hazelton}, {Sullivan},
  {Carroll}, {Barry}, {Rahimi}, {Pindor}, {Trott}, {Line}, {Jacobs}, {Morales},
  {Pober}, {Bernardi}, {Bowman}, {Busch}, {Briggs}, {Cappallo}, {Corey}, {de
  Oliveira-Costa}, {Dillon}, {Emrich}, {Ewall-Wice}, {Feng}, {Gaensler},
  {Goeke}, {Greenhill}, {Hewitt}, {Hurley-Walker}, {Johnston-Hollitt},
  {Kaplan}, {Kasper}, {Kim}, {Kratzenberg}, {Lenc}, {Loeb}, {Lonsdale},
  {Lynch}, {McKinley}, {McWhirter}, {Mitchell}, {Morgan}, {Neben},
  {Thyagarajan}, {Oberoi}, {Offringa}, {Ord}, {Paul}, {Prabu}, {Procopio},
  {Riding}, {Rogers}, {Roshi}, {Udaya Shankar}, {Sethi}, {Srivani},
  {Subrahmanyan}, {Tegmark}, {Tingay}, {Waterson}, {Wayth}, {Webster},
  {Whitney}, {Williams}, {Williams}, {Wu}, \&
  {Wyithe}}]{Beardsley2016ApJ...833..102B}
{Beardsley} A.~P. {et~al.}, 2016, \apj, 833, 102

\bibitem[{{Behroozi} {et~al}\mbox{.}(2019){Behroozi}, {Wechsler}, {Hearin}, \&
  {Conroy}}]{Behroozi2019MNRAS.488.3143B}
{Behroozi} P., {Wechsler} R.~H., {Hearin} A.~P., {Conroy} C., 2019, \mnras,
  488, 3143

\bibitem[{{Bhatawdekar} {et~al}\mbox{.}(2019){Bhatawdekar}, {Conselice},
  {Margalef-Bentabol}, \& {Duncan}}]{Bhatawdekar2019MNRAS.tmp..843B}
{Bhatawdekar} R., {Conselice} C.~J., {Margalef-Bentabol} B., {Duncan} K., 2019,
  \mnras, 843

\bibitem[{{Bouwens} {et~al}\mbox{.}(2015){Bouwens}, {Illingworth}, {Oesch},
  {Trenti}, {Labb{\'e}}, {Bradley}, {Carollo}, {van Dokkum}, {Gonzalez},
  {Holwerda}, {Franx}, {Spitler}, {Smit}, \& {Magee}}]{Bouwens2015a}
{Bouwens} R.~J. {et~al.}, 2015, \apj, 803, 34

\bibitem[{{Bouwens} {et~al}\mbox{.}(2017){Bouwens}, {Oesch}, {Illingworth},
  {Ellis}, \& {Stefanon}}]{Bouwens2017ApJ...843..129B}
{Bouwens} R.~J., {Oesch} P.~A., {Illingworth} G.~D., {Ellis} R.~S., {Stefanon}
  M., 2017, \apj, 843, 129

\bibitem[{{Bouwens} {et~al}\mbox{.}(2016){Bouwens}, {Oesch}, {Labb{\'e}},
  {Illingworth}, {Fazio}, {Coe}, {Holwerda}, {Smit}, {Stefanon}, {van Dokkum},
  {Trenti}, {Ashby}, {Huang}, {Spitler}, {Straatman}, {Bradley}, \&
  {Magee}}]{Bouwens2016}
{Bouwens} R.~J. {et~al.}, 2016, \apj, 830, 67

\bibitem[{Bowman {et~al}\mbox{.}(2018)Bowman, Rogers, Monsalve, Mozdzen, \&
  Mahesh}]{Bowman2018}
Bowman J.~D., Rogers A.~E., Monsalve R.~A., Mozdzen T.~J., Mahesh N., 2018,
  Nature, 555, 67

\bibitem[{{Bradley} {et~al}\mbox{.}(2019){Bradley}, {Tauscher}, {Rapetti}, \&
  {Burns}}]{Bradley2019ApJ...874..153B}
{Bradley} R.~F., {Tauscher} K., {Rapetti} D., {Burns} J.~O., 2019, \apj, 874,
  153

\bibitem[{{Bromm} {et~al}\mbox{.}(2001){Bromm}, {Kudritzki}, \&
  {Loeb}}]{Bromm2001ApJ...552..464B}
{Bromm} V., {Kudritzki} R.~P., {Loeb} A., 2001, \apj, 552, 464

\bibitem[{{Bromm} \& {Larson}(2004)}]{Bromm2004ARA&A..42...79B}
{Bromm} V., {Larson} R.~B., 2004, \araa, 42, 79

\bibitem[{{Brorby} {et~al}\mbox{.}(2016){Brorby}, {Kaaret}, {Prestwich}, \&
  {Mirabel}}]{Brorby2016MNRAS.457.4081B}
{Brorby} M., {Kaaret} P., {Prestwich} A., {Mirabel} I.~F., 2016, \mnras, 457,
  4081

\bibitem[{{Cen}(2003)}]{Cen2003ApJ...591...12C}
{Cen} R., 2003, \apj, 591, 12

\bibitem[{{Ciardi} {et~al}\mbox{.}(2006){Ciardi}, {Scannapieco}, {Stoehr},
  {Ferrara}, {Iliev}, \& {Shapiro}}]{Ciardi2006MNRAS.366..689C}
{Ciardi} B., {Scannapieco} E., {Stoehr} F., {Ferrara} A., {Iliev} I.~T.,
  {Shapiro} P.~R., 2006, \mnras, 366, 689

\bibitem[{Dalal {et~al}\mbox{.}(2010)Dalal, Pen, \& Seljak}]{Dalal:2010yt}
Dalal N., Pen U.-L., Seljak U., 2010, JCAP, 1011, 007

\bibitem[{{Dalla Vecchia} \& Schaye(2008)}]{DallaVecchia2008}
{Dalla Vecchia} C., Schaye J., 2008, \mnras, 387, 1431

\bibitem[{{Dalla Vecchia} \& Schaye(2012)}]{DallaVecchia2012}
{Dalla Vecchia} C., Schaye J., 2012, \mnras, 426, 140

\bibitem[{Das {et~al}\mbox{.}(2017)Das, Mesinger, Pallottini, Ferrara, \&
  Wise}]{Das_2017}
Das A., Mesinger A., Pallottini A., Ferrara A., Wise J.~H., 2017, \mnras, 469,
  1166–1174

\bibitem[{{Datta} {et~al}\mbox{.}(2014){Datta}, {Jensen}, {Majumdar},
  {Mellema}, {Iliev}, {Mao}, {Shapiro}, \& {Ahn}}]{Datta2014MNRAS.442.1491D}
{Datta} K.~K., {Jensen} H., {Majumdar} S., {Mellema} G., {Iliev} I.~T., {Mao}
  Y., {Shapiro} P.~R., {Ahn} K., 2014, \mnras, 442, 1491

\bibitem[{{Datta} {et~al}\mbox{.}(2012){Datta}, {Mellema}, {Mao}, {Iliev},
  {Shapiro}, \& {Ahn}}]{Datta2012MNRAS.424.1877D}
{Datta} K.~K., {Mellema} G., {Mao} Y., {Iliev} I.~T., {Shapiro} P.~R., {Ahn}
  K., 2012, \mnras, 424, 1877

\bibitem[{{Dayal} {et~al}\mbox{.}(2014){Dayal}, {Ferrara}, {Dunlop}, \&
  {Pacucci}}]{Dayal2014MNRAS.445.2545D}
{Dayal} P., {Ferrara} A., {Dunlop} J.~S., {Pacucci} F., 2014, \mnras, 445, 2545

\bibitem[{{DeBoer} {et~al}\mbox{.}(2017){DeBoer}, {Parsons}, {Aguirre},
  {Alexander}, {Ali}, {Beardsley}, {Bernardi}, {Bowman}, {Bradley}, {Carilli},
  {Cheng}, {de Lera Acedo}, {Dillon}, {Ewall-Wice}, {Fadana}, {Fagnoni},
  {Fritz}, {Furlanetto}, {Glendenning}, {Greig}, {Grobbelaar}, {Hazelton},
  {Hewitt}, {Hickish}, {Jacobs}, {Julius}, {Kariseb}, {Kohn}, {Lekalake},
  {Liu}, {Loots}, {MacMahon}, {Malan}, {Malgas}, {Maree}, {Martinot},
  {Mathison}, {Matsetela}, {Mesinger}, {Morales}, {Neben}, {Patra}, {Pieterse},
  {Pober}, {Razavi-Ghods}, {Ringuette}, {Robnett}, {Rosie}, {Sell}, {Smith},
  {Syce}, {Tegmark}, {Thyagarajan}, {Williams}, \&
  {Zheng}}]{DeBoer2017PASP..129d5001D}
{DeBoer} D.~R. {et~al.}, 2017, \pasp, 129, 045001

\bibitem[{{Dijkstra} {et~al}\mbox{.}(2004){Dijkstra}, {Haiman}, {Rees}, \&
  {Weinberg}}]{Dijkstra2004ApJ...601..666D}
{Dijkstra} M., {Haiman} Z., {Rees} M.~J., {Weinberg} D.~H., 2004, \apj, 601,
  666

\bibitem[{{Douna} {et~al}\mbox{.}(2015){Douna}, {Pellizza}, {Mirabel}, \&
  {Pedrosa}}]{Douna2015A&A...579A..44D}
{Douna} V.~M., {Pellizza} L.~J., {Mirabel} I.~F., {Pedrosa} S.~E., 2015, \aap,
  579, A44

\bibitem[{{Draine} \& {Bertoldi}(1996)}]{Draine1996ApJ...468..269D}
{Draine} B.~T., {Bertoldi} F., 1996, \apj, 468, 269

\bibitem[{{Efstathiou}(1992)}]{Efstathiou1992MNRAS.256P..43E}
{Efstathiou} G., 1992, \mnras, 256, 43P

\bibitem[{{Eldridge} {et~al}\mbox{.}(2017){Eldridge}, {Stanway}, {Xiao},
  {McClelland }, {Taylor}, {Ng}, {Greis}, \&
  {Bray}}]{Eldridge2017PASA...34...58E}
{Eldridge} J.~J., {Stanway} E.~R., {Xiao} L., {McClelland } L.~A.~S., {Taylor}
  G., {Ng} M., {Greis} S.~M.~L., {Bray} J.~C., 2017, \pasa, 34, e058

\bibitem[{{Evoli} {et~al}\mbox{.}(2014){Evoli}, {Mesinger}, \&
  {Ferrara}}]{Evoli2014JCAP...11..024E}
{Evoli} C., {Mesinger} A., {Ferrara} A., 2014, \jcap, 2014, 024

\bibitem[{{Ewall-Wice} {et~al}\mbox{.}(2018){Ewall-Wice}, {Chang}, {Lazio},
  {Dor{\'e}}, {Seiffert}, \& {Monsalve}}]{Ewall-Wice2018ApJ...868...63E}
{Ewall-Wice} A., {Chang} T.~C., {Lazio} J., {Dor{\'e}} O., {Seiffert} M.,
  {Monsalve} R.~A., 2018, \apj, 868, 63

\bibitem[{{Ferrara} \& {Loeb}(2013)}]{Ferrara2013MNRAS.431.2826F}
{Ferrara} A., {Loeb} A., 2013, \mnras, 431, 2826

\bibitem[{{Fialkov} \& {Barkana}(2014)}]{Fialkov2014MNRAS.445..213F}
{Fialkov} A., {Barkana} R., 2014, \mnras, 445, 213

\bibitem[{{Fialkov} {et~al}\mbox{.}(2018){Fialkov}, {Barkana}, \&
  {Cohen}}]{Fialkov2018PhRvL.121a1101F}
{Fialkov} A., {Barkana} R., {Cohen} A., 2018, Physical Review Letters, 121,
  011101

\bibitem[{{Fialkov} {et~al}\mbox{.}(2012){Fialkov}, {Barkana}, {Tseliakhovich},
  \& {Hirata}}]{Fialkov2012MNRAS.424.1335F}
{Fialkov} A., {Barkana} R., {Tseliakhovich} D., {Hirata} C.~M., 2012, \mnras,
  424, 1335

\bibitem[{{Fialkov} {et~al}\mbox{.}(2013){Fialkov}, {Barkana}, {Visbal},
  {Tseliakhovich}, \& {Hirata}}]{Fialkov2013MNRAS.432.2909F}
{Fialkov} A., {Barkana} R., {Visbal} E., {Tseliakhovich} D., {Hirata} C.~M.,
  2013, \mnras, 432, 2909

\bibitem[{{Finkelstein}(2016)}]{Finkelstein2016PASA...33...37F}
{Finkelstein} S.~L., 2016, \pasa, 33, e037

\bibitem[{{Finkelstein} {et~al}\mbox{.}(2015){Finkelstein}, {Ryan}, {Papovich},
  {Dickinson}, {Song}, {Somerville}, {Ferguson}, {Salmon}, {Giavalisco},
  {Koekemoer}, {Ashby}, {Behroozi}, {Castellano}, {Dunlop}, {Faber}, {Fazio},
  {Fontana}, {Grogin}, {Hathi}, {Jaacks}, {Kocevski}, {Livermore}, {McLure},
  {Merlin}, {Mobasher}, {Newman}, {Rafelski}, {Tilvi}, \&
  {Willner}}]{Finkelstein2015ApJ...810...71F}
{Finkelstein} S.~L. {et~al.}, 2015, \apj, 810, 71

\bibitem[{{Foreman-Mackey} {et~al}\mbox{.}(2013){Foreman-Mackey}, {Hogg},
  {Lang}, \& {Goodman}}]{Foreman2013PASP..125..306F}
{Foreman-Mackey} D., {Hogg} D.~W., {Lang} D., {Goodman} J., 2013, \pasp, 125,
  306

\bibitem[{{Fragos} {et~al}\mbox{.}(2013){Fragos}, {Lehmer}, {Tremmel},
  {Tzanavaris}, {Basu-Zych}, {Belczynski}, {Hornschemeier}, {Jenkins},
  {Kalogera}, {Ptak}, \& {Zezas}}]{Fragos2013ApJ...764...41F}
{Fragos} T. {et~al.}, 2013, \apj, 764, 41

\bibitem[{{Furlanetto} \& {Furlanetto}(2007)}]{Furlanetto2007MNRAS.374..547F}
{Furlanetto} S.~R., {Furlanetto} M.~R., 2007, \mnras, 374, 547

\bibitem[{Furlanetto \& Oh(2016)}]{furlanetto2016reionization}
Furlanetto S.~R., Oh S.~P., 2016, \mnras, 457, 1813

\bibitem[{{Furlanetto} {et~al}\mbox{.}(2006){Furlanetto}, {Oh}, \&
  {Briggs}}]{Furlanetto2006PhR...433..181F}
{Furlanetto} S.~R., {Oh} S.~P., {Briggs} F.~H., 2006, \physrep, 433, 181

\bibitem[{{Furlanetto} \& {Stoever}(2010)}]{Furlanetto2010MNRAS.404.1869F}
{Furlanetto} S.~R., {Stoever} S.~J., 2010, \mnras, 404, 1869

\bibitem[{{Furlanetto} {et~al}\mbox{.}(2004){Furlanetto}, {Zaldarriaga}, \&
  {Hernquist}}]{Furlanetto2004ApJ...613....1F}
{Furlanetto} S.~R., {Zaldarriaga} M., {Hernquist} L., 2004, \apj, 613, 1

\bibitem[{{Garaldi} {et~al}\mbox{.}(2019){Garaldi}, {Compostella}, \&
  {Porciani}}]{Garaldi2019MNRAS.483.5301G}
{Garaldi} E., {Compostella} M., {Porciani} C., 2019, \mnras, 483, 5301

\bibitem[{{Gardner} {et~al}\mbox{.}(2006){Gardner}, {Mather}, {Clampin},
  {Doyon}, {Greenhouse}, {Hammel}, {Hutchings}, {Jakobsen}, {Lilly}, {Long},
  {Lunine}, {McCaughrean}, {Mountain}, {Nella}, {Rieke}, {Rieke}, {Rix},
  {Smith}, {Sonneborn}, {Stiavelli}, {Stockman}, {Windhorst}, \&
  {Wright}}]{Gardner2006SSRv..123..485G}
{Gardner} J.~P. {et~al.}, 2006, \ssr, 123, 485

\bibitem[{{Goodman} \& {Weare}(2010)}]{Goodman2010CAMCS...5...65G}
{Goodman} J., {Weare} J., 2010, Communications in Applied Mathematics and
  Computational Science, 5, 65

\bibitem[{{Greif} {et~al}\mbox{.}(2011){Greif}, {White}, {Klessen}, \&
  {Springel}}]{Greif2011ApJ...736..147G}
{Greif} T.~H., {White} S. D.~M., {Klessen} R.~S., {Springel} V., 2011, \apj,
  736, 147

\bibitem[{{Greig} \& {Mesinger}(2015)}]{Greig2015MNRAS.449.4246G}
{Greig} B., {Mesinger} A., 2015, \mnras, 449, 4246

\bibitem[{{Greig} \& {Mesinger}(2017)}]{Greig2017MNRAS.472.2651G}
{Greig} B., {Mesinger} A., 2017, \mnras, 472, 2651

\bibitem[{Greig \& Mesinger(2017)}]{Greig2016}
Greig B., Mesinger A., 2017, \mnras, 465, 4838

\bibitem[{{Greig} \& {Mesinger}(2018)}]{Greig2018MNRAS.477.3217G}
{Greig} B., {Mesinger} A., 2018, \mnras, 477, 3217

\bibitem[{{Haiman} {et~al}\mbox{.}(2000){Haiman}, {Abel}, \&
  {Rees}}]{Haiman2000ApJ...534...11H}
{Haiman} Z., {Abel} T., {Rees} M.~J., 2000, \apj, 534, 11

\bibitem[{{Haiman} {et~al}\mbox{.}(1996){Haiman}, {Rees}, \&
  {Loeb}}]{Haiman1996ApJ...467..522H}
{Haiman} Z., {Rees} M.~J., {Loeb} A., 1996, \apj, 467, 522

\bibitem[{{Haiman} {et~al}\mbox{.}(1997){Haiman}, {Rees}, \&
  {Loeb}}]{Haiman1997ApJ...476..458H}
{Haiman} Z., {Rees} M.~J., {Loeb} A., 1997, \apj, 476, 458

\bibitem[{Hassan {et~al}\mbox{.}(2016)Hassan, Dav{\'{e}}, Finlator, \&
  Santos}]{Hassan2016}
Hassan S., Dav{\'{e}} R., Finlator K., Santos M.~G., 2016, \mnras, 457, 1550

\bibitem[{{Hektor} {et~al}\mbox{.}(2018){Hektor}, {H{\"u}tsi}, {Marzola},
  {Raidal}, {Vaskonen}, \& {Veerm{\"a}e}}]{Hektor2018PhRvD..98b3503H}
{Hektor} A., {H{\"u}tsi} G., {Marzola} L., {Raidal} M., {Vaskonen} V.,
  {Veerm{\"a}e} H., 2018, \prd, 98, 023503

\bibitem[{Hills {et~al}\mbox{.}(2018)Hills, Kulkarni, Meerburg, \&
  Puchwein}]{Hills2018arXiv180501421H}
Hills R., Kulkarni G., Meerburg P.~D., Puchwein E., 2018, Nature, 564,
  E32–E34

\bibitem[{{Hirata}(2006)}]{Hirata2006MNRAS.367..259H}
{Hirata} C.~M., 2006, \mnras, 367, 259

\bibitem[{{Holzbauer} \& {Furlanetto}(2012)}]{Holzbauer2012MNRAS.419..718H}
{Holzbauer} L.~N., {Furlanetto} S.~R., 2012, \mnras, 419, 718

\bibitem[{{Hopkins} {et~al}\mbox{.}(2014){Hopkins}, {Kere{\v{s}}},
  {O{\~n}orbe}, {Faucher-Gigu{\`e}re}, {Quataert}, {Murray}, \&
  {Bullock}}]{Hopkins2014MNRAS.445..581H}
{Hopkins} P.~F., {Kere{\v{s}}} D., {O{\~n}orbe} J., {Faucher-Gigu{\`e}re}
  C.-A., {Quataert} E., {Murray} N., {Bullock} J.~S., 2014, \mnras, 445, 581

\bibitem[{Hopkins {et~al}\mbox{.}(2018)Hopkins, Wetzel, Kereš,
  Faucher-Giguère, Quataert, Boylan-Kolchin, Murray, Hayward, Garrison-Kimmel,
  Hummels, \& et~al.}]{Hopkins2017}
Hopkins P.~F. {et~al.}, 2018, \mnras, 480, 800–863

\bibitem[{{Hui} \& {Gnedin}(1997)}]{Hui1997MNRAS.292...27H}
{Hui} L., {Gnedin} N.~Y., 1997, \mnras, 292, 27

\bibitem[{{Ishigaki} {et~al}\mbox{.}(2018){Ishigaki}, {Kawamata}, {Ouchi},
  {Oguri}, {Shimasaku}, \& {Ono}}]{Ishigaki2018ApJ...854...73I}
{Ishigaki} M., {Kawamata} R., {Ouchi} M., {Oguri} M., {Shimasaku} K., {Ono} Y.,
  2018, \apj, 854, 73

\bibitem[{Jaacks {et~al}\mbox{.}(2019)Jaacks, Finkelstein, \&
  Bromm}]{Jaacks2018}
Jaacks J., Finkelstein S.~L., Bromm V., 2019, \mnras, 488, 2202–2221

\bibitem[{{Jensen} {et~al}\mbox{.}(2013){Jensen}, {Datta}, {Mellema},
  {Chapman}, {Abdalla}, {Iliev}, {Mao}, {Santos}, {Shapiro}, {Zaroubi},
  {Bernardi}, {Brentjens}, {de Bruyn}, {Ciardi}, {Harker}, {Jeli{\'c}},
  {Kazemi}, {Koopmans}, {Labropoulos}, {Martinez}, {Offringa}, {Pand ey},
  {Schaye}, {Thomas}, {Veligatla}, {Vedantham}, \&
  {Yatawatta}}]{Jensen2013MNRAS.435..460J}
{Jensen} H. {et~al.}, 2013, \mnras, 435, 460

\bibitem[{{Johnson} {et~al}\mbox{.}(2007){Johnson}, {Greif}, \&
  {Bromm}}]{Johnson2007ApJ...665...85J}
{Johnson} J.~L., {Greif} T.~H., {Bromm} V., 2007, \apj, 665, 85

\bibitem[{Katz {et~al}\mbox{.}(2020)Katz, Ramsoy, Rosdahl, Kimm, Blaizot,
  Haehnelt, Michel-Dansac, Garel, Laigle, Devriendt, \&
  et~al.}]{Katz2019arXiv190511414K}
Katz H. {et~al.}, 2020, \mnras, 494, 2200–2220

\bibitem[{Keller {et~al}\mbox{.}(2014)Keller, Wadsley, Benincasa, \&
  Couchman}]{Keller2014}
Keller B.~W., Wadsley J., Benincasa S.~M., Couchman H. M.~P., 2014, \mnras,
  442, 3013

\bibitem[{Kimm \& Cen(2014)}]{Kimm2014}
Kimm T., Cen R., 2014, \apj, 788, 121

\bibitem[{Kimm {et~al}\mbox{.}(2017)Kimm, Katz, Haehnelt, Rosdahl, Devriendt,
  \& Slyz}]{Kimm2016}
Kimm T., Katz H., Haehnelt M., Rosdahl J., Devriendt J., Slyz A., 2017, \mnras,
  stx052

\bibitem[{{Koh} \& {Wise}(2018)}]{Koh2018MNRAS.474.3817K}
{Koh} D., {Wise} J.~H., 2018, \mnras, 474, 3817

\bibitem[{{Kohn} {et~al}\mbox{.}(2019){Kohn}, {Aguirre}, {La Plante},
  {Billings}, {Chichura}, {Fortino}, {Igarashi}, {Benefo}, {Gallardo},
  {Martinot}, {Nunhokee}, {Kern}, {Bull}, {Liu}, {Alexander}, {Ali},
  {Beardsley}, {Bernardi}, {Bowman}, {Bradley}, {Carilli}, {Cheng}, {DeBoer},
  {de Lera Acedo}, {Dillon}, {Ewall-Wice}, {Fadana}, {Fagnoni}, {Fritz},
  {Furlanetto}, {Glendenning}, {Greig}, {Grobbelaar}, {Hazelton}, {Hewitt},
  {Hickish}, {Jacobs}, {Julius}, {Kariseb}, {Kolopanis}, {Lekalake}, {Loots},
  {MacMahon}, {Malan}, {Malgas}, {Maree}, {Mathison}, {Matsetela}, {Mesinger},
  {Morales}, {Neben}, {Nikolic}, {Parsons}, {Patra}, {Pieterse}, {Pober},
  {Razavi-Ghods}, {Ringuette}, {Robnett}, {Rosie}, {Sell}, {Smith}, {Syce},
  {Tegmark}, {Thyagarajan}, {Williams}, \& {Zheng}}]{Kohn2019ApJ...882...58K}
{Kohn} S.~A. {et~al.}, 2019, \apj, 882, 58

\bibitem[{Koopmans {et~al}\mbox{.}(2015)Koopmans, Pritchard, Mellema, Abdalla,
  Aguirre, Ahn, Barkana, van Bemmel, Bernardi, Bonaldi, Briggs, de~Bruyn,
  Chang, Chapman, Chen, Ciardi, Datta, Dayal, Ferrara, Fialkov, Fiore, Ichiki,
  Illiev, Inoue, Jelić, Jones, Lazio, Maio, Majumdar, Mack, Mesinger, Morales,
  Parsons, Pen, Santos, Schneider, Semelin, de~Souza, Subrahmanyan, Takeuchi,
  Trott, Vedantham, Wagg, Webster, \& Wyithe}]{Koopmans2015aska.confE...1K}
Koopmans L. V.~E. {et~al.}, 2015

\bibitem[{Kuhlen \& Faucher-Gigu{\`{e}}re(2012)}]{Kuhlen2012}
Kuhlen M., Faucher-Gigu{\`{e}}re C.-A., 2012, \mnras, 423, 862

\bibitem[{{Leitherer} {et~al}\mbox{.}(1999){Leitherer}, {Schaerer}, {Goldader},
  {Delgado}, {Robert}, {Kune}, {de Mello}, {Devost}, \&
  {Heckman}}]{Leitherer1999ApJS..123....3L}
{Leitherer} C. {et~al.}, 1999, \apjs, 123, 3

\bibitem[{{Lippai} {et~al}\mbox{.}(2009){Lippai}, {Frei}, \&
  {Haiman}}]{Lippai2009ApJ...701..360L}
{Lippai} Z., {Frei} Z., {Haiman} Z., 2009, \apj, 701, 360

\bibitem[{Liu {et~al}\mbox{.}(2016)Liu, Mutch, Angel, Duffy, Geil, Poole,
  Mesinger, \& Wyithe}]{Liu2016}
Liu C., Mutch S.~J., Angel P.~W., Duffy A.~R., Geil P.~M., Poole G.~B.,
  Mesinger A., Wyithe J. S.~B., 2016, \mnras, 462, 235

\bibitem[{{Livermore} {et~al}\mbox{.}(2017){Livermore}, {Finkelstein}, \&
  {Lotz}}]{Livermore2017ApJ...835..113L}
{Livermore} R.~C., {Finkelstein} S.~L., {Lotz} J.~M., 2017, \apj, 835, 113

\bibitem[{{Lopez-Honorez} {et~al}\mbox{.}(2016){Lopez-Honorez}, {Mena},
  {Molin{\'e}}, {Palomares-Ruiz}, \&
  {Vincent}}]{Lopez-Honorez2016JCAP...08..004L}
{Lopez-Honorez} L., {Mena} O., {Molin{\'e}} {\'A}., {Palomares-Ruiz} S.,
  {Vincent} A.~C., 2016, \jcap, 2016, 004

\bibitem[{Ma {et~al}\mbox{.}(2018)Ma, Hopkins, Garrison-Kimmel,
  Faucher-Gigu{\`{e}}re, Quataert, Boylan-Kolchin, Hayward, Feldmann, \&
  Kere{\v{s}}}]{Ma2018}
Ma X. {et~al.}, 2018, \mnras, 478, 1694

\bibitem[{{Ma} {et~al}\mbox{.}(2015){Ma}, {Kasen}, {Hopkins},
  {Faucher-Gigu{\`e}re}, {Quataert}, {Kere{\v{s}}}, \&
  {Murray}}]{Ma2015MNRAS.453..960M}
{Ma} X., {Kasen} D., {Hopkins} P.~F., {Faucher-Gigu{\`e}re} C.-A., {Quataert}
  E., {Kere{\v{s}}} D., {Murray} N., 2015, \mnras, 453, 960

\bibitem[{Madau(2018)}]{Madau2018}
Madau P., 2018, Monthly Notices of the Royal Astronomical Society: Letters,
  480, L43–L47

\bibitem[{{Madau} \& {Dickinson}(2014)}]{Madau2014ARA&A..52..415M}
{Madau} P., {Dickinson} M., 2014, Annual Review of Astronomy and Astrophysics,
  52, 415

\bibitem[{Manti {et~al}\mbox{.}(2017)Manti, Gallerani, Ferrara, Greig, \&
  Feruglio}]{Manti2017}
Manti S., Gallerani S., Ferrara A., Greig B., Feruglio C., 2017, \mnras, 466,
  1160

\bibitem[{{Mao} {et~al}\mbox{.}(2012){Mao}, {Shapiro}, {Mellema}, {Iliev},
  {Koda}, \& {Ahn}}]{Mao2012MNRAS.422..926M}
{Mao} Y., {Shapiro} P.~R., {Mellema} G., {Iliev} I.~T., {Koda} J., {Ahn} K.,
  2012, \mnras, 422, 926

\bibitem[{{Mapelli} {et~al}\mbox{.}(2010){Mapelli}, {Ripamonti}, {Zampieri},
  {Colpi}, \& {Bressan}}]{Mapelli2010MNRAS.408..234M}
{Mapelli} M., {Ripamonti} E., {Zampieri} L., {Colpi} M., {Bressan} A., 2010,
  \mnras, 408, 234

\bibitem[{Mason {et~al}\mbox{.}(2015)Mason, Trenti, \& Treu}]{Mason2015}
Mason C.~A., Trenti M., Treu T., 2015, \apj, 813, 21

\bibitem[{{McGreer} {et~al}\mbox{.}(2015){McGreer}, {Mesinger}, \&
  {D'Odorico}}]{McGreer2015MNRAS.447..499M}
{McGreer} I.~D., {Mesinger} A., {D'Odorico} V., 2015, \mnras, 447, 499

\bibitem[{McQuinn \& D’Aloisio(2018)}]{mcquinn2018observable}
McQuinn M., D’Aloisio A., 2018, Journal of Cosmology and Astroparticle
  Physics, 2018, 016–016

\bibitem[{{McQuinn} {et~al}\mbox{.}(2007){McQuinn}, {Lidz}, {Zahn}, {Dutta},
  {Hernquist}, \& {Zaldarriaga}}]{McQuinn2007MNRAS.377.1043M}
{McQuinn} M., {Lidz} A., {Zahn} O., {Dutta} S., {Hernquist} L., {Zaldarriaga}
  M., 2007, \mnras, 377, 1043

\bibitem[{{McQuinn} \& {O'Leary}(2012)}]{McQuinn2012ApJ...760....3M}
{McQuinn} M., {O'Leary} R.~M., 2012, \apj, 760, 3

\bibitem[{Mebane {et~al}\mbox{.}(2020)Mebane, Mirocha, \&
  Furlanetto}]{Mebane2019arXiv191010171M}
Mebane R.~H., Mirocha J., Furlanetto S.~R., 2020, \mnras, 493, 1217–1226

\bibitem[{{Mellema} {et~al}\mbox{.}(2013){Mellema}, {Koopmans}, {Abdalla},
  {Bernardi}, {Ciardi}, {Daiboo}, {de Bruyn}, {Datta}, {Falcke}, {Ferrara},
  {Iliev}, {Iocco}, {Jeli{\'c}}, {Jensen}, {Joseph}, {Labroupoulos}, {Meiksin},
  {Mesinger}, {Offringa}, {Pandey}, {Pritchard}, {Santos}, {Schwarz},
  {Semelin}, {Vedantham}, {Yatawatta}, \&
  {Zaroubi}}]{Mellema2013ExA....36..235M}
{Mellema} G. {et~al.}, 2013, Experimental Astronomy, 36, 235

\bibitem[{{Mesinger} \& {Dijkstra}(2008)}]{Mesinger2008MNRAS.390.1071M}
{Mesinger} A., {Dijkstra} M., 2008, \mnras, 390, 1071

\bibitem[{{Mesinger} {et~al}\mbox{.}(2013){Mesinger}, {Ferrara}, \&
  {Spiegel}}]{Mesinger2013MNRAS.431..621M}
{Mesinger} A., {Ferrara} A., {Spiegel} D.~S., 2013, \mnras, 431, 621

\bibitem[{{Mesinger} \& {Furlanetto}(2007)}]{Mesinger2007ApJ...669..663M}
{Mesinger} A., {Furlanetto} S., 2007, \apj, 669, 663

\bibitem[{{Mesinger} {et~al}\mbox{.}(2011){Mesinger}, {Furlanetto}, \&
  {Cen}}]{Mesinger2011MNRAS.411..955M}
{Mesinger} A., {Furlanetto} S., {Cen} R., 2011, \mnras, 411, 955

\bibitem[{Mesinger {et~al}\mbox{.}(2016)Mesinger, Greig, \&
  Sobacchi}]{Mesinger2016}
Mesinger A., Greig B., Sobacchi E., 2016, \mnras, 459, 2342

\bibitem[{{Mesinger} {et~al}\mbox{.}(2012){Mesinger}, {McQuinn}, \&
  {Spergel}}]{Mesinger2012MNRAS.422.1403M}
{Mesinger} A., {McQuinn} M., {Spergel} D.~N., 2012, \mnras, 422, 1403

\bibitem[{Miller {et~al}\mbox{.}(2015)Miller, Gallo, Greene, Kelly, Treu, Woo,
  \& Baldassare}]{Miller_2015}
Miller B.~P., Gallo E., Greene J.~E., Kelly B.~C., Treu T., Woo J.-H.,
  Baldassare V., 2015, \apj, 799, 98

\bibitem[{Mineo {et~al}\mbox{.}(2012)Mineo, Gilfanov, \& Sunyaev}]{Mineo2012}
Mineo S., Gilfanov M., Sunyaev R., 2012, \mnras, 419, 2095

\bibitem[{{Miralda-Escud{\'e}} {et~al}\mbox{.}(2000){Miralda-Escud{\'e}},
  {Haehnelt}, \& {Rees}}]{Miralda2000ApJ...530....1M}
{Miralda-Escud{\'e}} J., {Haehnelt} M., {Rees} M.~J., 2000, \apj, 530, 1

\bibitem[{{Miranda} {et~al}\mbox{.}(2017){Miranda}, {Lidz}, {Heinrich}, \&
  {Hu}}]{Miranda2017MNRAS.467.4050M}
{Miranda} V., {Lidz} A., {Heinrich} C.~H., {Hu} W., 2017, \mnras, 467, 4050

\bibitem[{{Mirocha} \& {Furlanetto}(2019)}]{Mirocha2019MNRAS.483.1980M}
{Mirocha} J., {Furlanetto} S.~R., 2019, \mnras, 483, 1980

\bibitem[{Mirocha {et~al}\mbox{.}(2016)Mirocha, Furlanetto, \&
  Sun}]{Mirocha2016}
Mirocha J., Furlanetto S.~R., Sun G., 2016, \mnras, 464, 1365–1379

\bibitem[{Mirocha {et~al}\mbox{.}(2018)Mirocha, Mebane, Furlanetto, Singal, \&
  Trinh}]{Mirocha2017}
Mirocha J., Mebane R.~H., Furlanetto S.~R., Singal K., Trinh D., 2018, \mnras,
  478, 5591–5606

\bibitem[{Mitra {et~al}\mbox{.}(2015)Mitra, {Roy Choudhury}, \&
  Ferrara}]{Mitra2015}
Mitra S., {Roy Choudhury} T., Ferrara A., 2015, Monthly Notices of the Royal
  Astronomical Society: Letters, 454, L76

\bibitem[{{Moster} {et~al}\mbox{.}(2013){Moster}, {Naab}, \&
  {White}}]{Moster2013MNRAS.428.3121M}
{Moster} B.~P., {Naab} T., {White} S. D.~M., 2013, \mnras, 428, 3121

\bibitem[{{Mu{\~n}oz}(2019{\natexlab{a}})}]{Munoz2019PhRvD.100f3538M}
{Mu{\~n}oz} J.~B., 2019{\natexlab{a}}, \prd, 100, 063538

\bibitem[{{Mu{\~n}oz}(2019{\natexlab{b}})}]{Munoz:2019fkt}
{Mu{\~n}oz} J.~B., 2019{\natexlab{b}}, Phys. Rev. Lett., 123, 131301

\bibitem[{{Mu{\~n}oz} \& {Loeb}(2018)}]{Munoz2018Natur.557..684M}
{Mu{\~n}oz} J.~B., {Loeb} A., 2018, Nature, 557, 684

\bibitem[{Mutch {et~al}\mbox{.}(2016)Mutch, Geil, Poole, Angel, Duffy,
  Mesinger, \& Wyithe}]{Mutch2016}
Mutch S.~J., Geil P.~M., Poole G.~B., Angel P.~W., Duffy A.~R., Mesinger A.,
  Wyithe J. S.~B., 2016, \mnras, 462, 250

\bibitem[{Naidu {et~al}\mbox{.}(2020)Naidu, Tacchella, Mason, Bose, Oesch, \&
  Conroy}]{Naidu2019arXiv190713130N}
Naidu R.~P., Tacchella S., Mason C.~A., Bose S., Oesch P.~A., Conroy C., 2020,
  \apj, 892, 109

\bibitem[{{Noh} \& {McQuinn}(2014)}]{Noh2014MNRAS.444..503N}
{Noh} Y., {McQuinn} M., 2014, \mnras, 444, 503

\bibitem[{{Ocvirk} {et~al}\mbox{.}(2018){Ocvirk}, {Aubert}, {Sorce}, {Shapiro},
  {Deparis}, {Dawoodbhoy}, {Lewis}, {Teyssier}, {Yepes}, {Gottl{\"o}ber},
  {Ahn}, {Iliev}, \& {Hoffman}}]{Ocvirk2018arXiv181111192O}
{Ocvirk} P. {et~al.}, 2018, arXiv e-prints, arXiv:1811.11192

\bibitem[{{Oesch} {et~al}\mbox{.}(2018){Oesch}, {Bouwens}, {Illingworth},
  {Labb{\'e}}, \& {Stefanon}}]{Oesch2018ApJ...855..105O}
{Oesch} P.~A., {Bouwens} R.~J., {Illingworth} G.~D., {Labb{\'e}} I., {Stefanon}
  M., 2018, \apj, 855, 105

\bibitem[{{O'Leary} \& {McQuinn}(2012)}]{Oleary2012ApJ...760....4O}
{O'Leary} R.~M., {McQuinn} M., 2012, \apj, 760, 4

\bibitem[{{O'Shea} {et~al}\mbox{.}(2015){O'Shea}, {Wise}, {Xu}, \&
  {Norman}}]{OShea2015ApJ...807L..12O}
{O'Shea} B.~W., {Wise} J.~H., {Xu} H., {Norman} M.~L., 2015, \apjl, 807, L12

\bibitem[{{Paardekooper} {et~al}\mbox{.}(2015){Paardekooper}, {Khochfar}, \&
  {Dalla Vecchia}}]{Paardekooper2015MNRAS.451.2544P}
{Paardekooper} J.-P., {Khochfar} S., {Dalla Vecchia} C., 2015, \mnras, 451,
  2544

\bibitem[{Pacucci {et~al}\mbox{.}(2014)Pacucci, Mesinger, Mineo, \&
  Ferrara}]{Pacucci2014}
Pacucci F., Mesinger A., Mineo S., Ferrara A., 2014, \mnras, 443, 678–686

\bibitem[{{Park} {et~al}\mbox{.}(2020){Park}, {Gillet}, {Mesinger}, \&
  {Greig}}]{Park2020MNRAS.491.3891P}
{Park} J., {Gillet} N., {Mesinger} A., {Greig} B., 2020, \mnras, 491, 3891

\bibitem[{{Park} {et~al}\mbox{.}(2019){Park}, {Mesinger}, {Greig}, \&
  {Gillet}}]{Park2019MNRAS.484..933P}
{Park} J., {Mesinger} A., {Greig} B., {Gillet} N., 2019, \mnras, 484, 933

\bibitem[{Parsa {et~al}\mbox{.}(2017)Parsa, Dunlop, \& McLure}]{Parsa2017}
Parsa S., Dunlop J.~S., McLure R.~J., 2017, \mnras, 474, 2904–2923

\bibitem[{{Patil} {et~al}\mbox{.}(2017){Patil}, {Yatawatta}, {Koopmans}, {de
  Bruyn}, {Brentjens}, {Zaroubi}, {Asad}, {Hatef}, {Jeli{\'c}}, {Mevius},
  {Offringa}, {Pandey}, {Vedantham}, {Abdalla}, {Brouw}, {Chapman}, {Ciardi},
  {Gehlot}, {Ghosh}, {Harker}, {Iliev}, {Kakiichi}, {Majumdar}, {Mellema},
  {Silva}, {Schaye}, {Vrbanec}, \& {Wijnholds}}]{Patil2017ApJ...838...65P}
{Patil} A.~H. {et~al.}, 2017, \apj, 838, 65

\bibitem[{{Planck Collaboration} {et~al}\mbox{.}(2016{\natexlab{a}}){Planck
  Collaboration}, {Adam}, {Aghanim}, {Ashdown}, {Aumont}, {Baccigalupi},
  {Ballardini}, {Banday}, {Barreiro}, {Bartolo}, {Basak}, {Battye}, {Benabed},
  {Bernard}, {Bersanelli}, {Bielewicz}, {Bock}, {Bonaldi}, {Bonavera}, {Bond},
  {Borrill}, {Bouchet}, {Boulanger}, {Bucher}, {Burigana}, {Calabrese},
  {Cardoso}, {Carron}, {Chiang}, {Colombo}, {Combet}, {Comis}, {Couchot},
  {Coulais}, {Crill}, {Curto}, {Cuttaia}, {Davis}, {de Bernardis}, {de Rosa},
  {de Zotti}, {Delabrouille}, {Di Valentino}, {Dickinson}, {Diego}, {Dor{\'e}},
  {Douspis}, {Ducout}, {Dupac}, {Elsner}, {En{\ss}lin}, {Eriksen}, {Falgarone},
  {Fantaye}, {Finelli}, {Forastieri}, {Frailis}, {Fraisse}, {Franceschi},
  {Frolov}, {Galeotta}, {Galli}, {Ganga}, {G{\'e}nova-Santos}, {Gerbino},
  {Ghosh}, {Gonz{\'a}lez-Nuevo}, {G{\'o}rski}, {Gruppuso}, {Gudmundsson},
  {Hansen}, {Helou}, {Henrot-Versill{\'e}}, {Herranz}, {Hivon}, {Huang},
  {Ili{\'c}}, {Jaffe}, {Jones}, {Keih{\"a}nen}, {Keskitalo}, {Kisner}, {Knox},
  {Krachmalnicoff}, {Kunz}, {Kurki-Suonio}, {Lagache}, {L{\"a}hteenm{\"a}ki},
  {Lamarre}, {Langer}, {Lasenby}, {Lattanzi}, {Lawrence}, {Le Jeune},
  {Levrier}, {Lewis}, {Liguori}, {Lilje}, {L{\'o}pez-Caniego}, {Ma},
  {Mac{\'\i}as-P{\'e}rez}, {Maggio}, {Mangilli}, {Maris}, {Martin},
  {Mart{\'\i}nez-Gonz{\'a}lez}, {Matarrese}, {Mauri}, {McEwen}, {Meinhold},
  {Melchiorri}, {Mennella}, {Migliaccio}, {Miville-Desch{\^e}nes}, {Molinari},
  {Moneti}, {Montier}, {Morgante}, {Moss}, {Naselsky}, {Natoli}, {Oxborrow},
  {Pagano}, {Paoletti}, {Partridge}, {Patanchon}, {Patrizii}, {Perdereau},
  {Perotto}, {Pettorino}, {Piacentini}, {Plaszczynski}, {Polastri}, {Polenta},
  {Puget}, {Rachen}, {Racine}, {Reinecke}, {Remazeilles}, {Renzi}, {Rocha},
  {Rossetti}, {Roudier}, {Rubi{\~n}o-Mart{\'\i}n}, {Ruiz-Granados}, {Salvati},
  {Sandri}, {Savelainen}, {Scott}, {Sirri}, {Sunyaev}, {Suur-Uski}, {Tauber},
  {Tenti}, {Toffolatti}, {Tomasi}, {Tristram}, {Trombetti}, {Valiviita}, {Van
  Tent}, {Vielva}, {Villa}, {Vittorio}, {Wandelt}, {Wehus}, {White}, {Zacchei},
  \& {Zonca}}]{Planck2016A&A...596A.108P}
{Planck Collaboration} {et~al.}, 2016{\natexlab{a}}, \aap, 596, A108

\bibitem[{{Planck Collaboration} {et~al}\mbox{.}(2016{\natexlab{b}}){Planck
  Collaboration}, {Ade}, {Aghanim}, {Arnaud}, {Ashdown}, {Aumont},
  {Baccigalupi}, {Banday}, {Barreiro}, {Bartlett}, \&
  et~al.}]{Planck2016A&A...594A..13P}
{Planck Collaboration} {et~al.}, 2016{\natexlab{b}}, \aap, 594, A13

\bibitem[{{Pritchard} \& {Furlanetto}(2007)}]{Pritchard2007MNRAS.376.1680P}
{Pritchard} J.~R., {Furlanetto} S.~R., 2007, \mnras, 376, 1680

\bibitem[{Qin {et~al}\mbox{.}(2019)Qin, Duffy, Mutch, Poole, Mesinger, \&
  Wyithe}]{Qin_2019}
Qin Y., Duffy A.~R., Mutch S.~J., Poole G.~B., Mesinger A., Wyithe J. S.~B.,
  2019, \mnras, 487, 1946–1963

\bibitem[{Qin {et~al}\mbox{.}(2017)Qin, Mutch, Poole, Liu, Angel, Duffy, Geil,
  Mesinger, \& Wyithe}]{Qin2017a}
Qin Y. {et~al.}, 2017, \mnras, 472, 2009–2027

\bibitem[{{Rahmati} {et~al}\mbox{.}(2013){Rahmati}, {Pawlik},
  {Rai{\v{c}}evi{\'c}}, \& {Schaye}}]{Rahmati2013MNRAS.430.2427R}
{Rahmati} A., {Pawlik} A.~H., {Rai{\v{c}}evi{\'c}} M., {Schaye} J., 2013,
  \mnras, 430, 2427

\bibitem[{{Ricci} {et~al}\mbox{.}(2017){Ricci}, {Marchesi}, {Shankar}, {La
  Franca}, \& {Civano}}]{Ricci2017MNRAS.465.1915R}
{Ricci} F., {Marchesi} S., {Shankar} F., {La Franca} F., {Civano} F., 2017,
  \mnras, 465, 1915

\bibitem[{{Ricotti} {et~al}\mbox{.}(2001){Ricotti}, {Gnedin}, \&
  {Shull}}]{Ricotti2001ApJ...560..580R}
{Ricotti} M., {Gnedin} N.~Y., {Shull} J.~M., 2001, \apj, 560, 580

\bibitem[{{Ricotti} \& {Ostriker}(2004)}]{RO04}
{Ricotti} M., {Ostriker} J.~P., 2004, \mnras, 352, 547

\bibitem[{Sanderbeck {et~al}\mbox{.}(2018)Sanderbeck, McQuinn, D’Aloisio, \&
  Werk}]{Sanderbeck_2018}
Sanderbeck P. R.~U., McQuinn M., D’Aloisio A., Werk J.~K., 2018, \apj, 869,
  159

\bibitem[{{Schaerer}(2002)}]{Schaerer2002A&A...382...28S}
{Schaerer} D., 2002, \aap, 382, 28

\bibitem[{{Schauer} {et~al}\mbox{.}(2019){Schauer}, {Glover}, {Klessen}, \&
  {Ceverino}}]{Schauer2019MNRAS.484.3510S}
{Schauer} A. T.~P., {Glover} S. C.~O., {Klessen} R.~S., {Ceverino} D., 2019,
  \mnras, 484, 3510

\bibitem[{{Schauer} {et~al}\mbox{.}(2015){Schauer}, {Whalen}, {Glover}, \&
  {Klessen}}]{Schauer2015MNRAS.454.2441S}
{Schauer} A. T.~P., {Whalen} D.~J., {Glover} S. C.~O., {Klessen} R.~S., 2015,
  \mnras, 454, 2441

\bibitem[{{Scoccimarro}(1998)}]{Scoccimarro1998MNRAS.299.1097S}
{Scoccimarro} R., 1998, \mnras, 299, 1097

\bibitem[{{Seager} {et~al}\mbox{.}(1999){Seager}, {Sasselov}, \&
  {Scott}}]{Seager1999ApJ...523L...1S}
{Seager} S., {Sasselov} D.~D., {Scott} D., 1999, \apjl, 523, L1

\bibitem[{{Shang} {et~al}\mbox{.}(2010){Shang}, {Bryan}, \&
  {Haiman}}]{Shang2010MNRAS.402.1249S}
{Shang} C., {Bryan} G.~L., {Haiman} Z., 2010, \mnras, 402, 1249

\bibitem[{{Shapiro} {et~al}\mbox{.}(1994){Shapiro}, {Giroux}, \&
  {Babul}}]{Shapiro1994ApJ...427...25S}
{Shapiro} P.~R., {Giroux} M.~L., {Babul} A., 1994, \apj, 427, 25

\bibitem[{{Sims} \& {Pober}(2019)}]{Sims2019MNRAS.488.2904S}
{Sims} P.~H., {Pober} J.~C., 2019, \mnras, 488, 2904

\bibitem[{{Sobacchi} \& {Mesinger}(2014)}]{Sobacchi2014MNRAS.440.1662S}
{Sobacchi} E., {Mesinger} A., 2014, \mnras, 440, 1662

\bibitem[{Sun \& Furlanetto(2016)}]{Sun2015}
Sun G., Furlanetto S.~R., 2016, \mnras, 460, 417

\bibitem[{{Tacchella} {et~al}\mbox{.}(2018){Tacchella}, {Bose}, {Conroy},
  {Eisenstein}, \& {Johnson}}]{Tacchella2018ApJ...868...92T}
{Tacchella} S., {Bose} S., {Conroy} C., {Eisenstein} D.~J., {Johnson} B.~D.,
  2018, \apj, 868, 92

\bibitem[{{Thoul} \& {Weinberg}(1996)}]{Thoul1996ApJ...465..608T}
{Thoul} A.~A., {Weinberg} D.~H., 1996, \apj, 465, 608

\bibitem[{Tingay {et~al}\mbox{.}(2013)Tingay, Goeke, Bowman, Emrich, Ord,
  Mitchell, Morales, Booler, Crosse, Wayth, \&
  et~al.}]{Tingay2013PASA...30....7T}
Tingay S.~J. {et~al.}, 2013, \pasa, 30

\bibitem[{{Trac} \& {Gnedin}(2011)}]{Trac2011ASL.....4..228T}
{Trac} H.~Y., {Gnedin} N.~Y., 2011, Advanced Science Letters, 4, 228

\bibitem[{{Tseliakhovich} \& {Hirata}(2010)}]{Tseliakhovich2010PhRvD..82h3520T}
{Tseliakhovich} D., {Hirata} C., 2010, \prd, 82, 083520

\bibitem[{{Tumlinson} \& {Shull}(2000)}]{Tumlinson2000ApJ...528L..65T}
{Tumlinson} J., {Shull} J.~M., 2000, \apj, 528, L65

\bibitem[{{van Haarlem} {et~al}\mbox{.}(2013){van Haarlem}, {Wise}, {Gunst},
  {Heald}, {McKean}, {Hessels}, {de Bruyn}, {Nijboer}, {Swinbank}, {Fallows},
  {Brentjens}, {Nelles}, {Beck}, {Falcke}, {Fender}, {H{\"o}randel},
  {Koopmans}, {Mann}, {Miley}, {R{\"o}ttgering}, {Stappers}, {Wijers},
  {Zaroubi}, {van den Akker}, {Alexov}, {Anderson}, {Anderson}, {van Ardenne},
  {Arts}, {Asgekar}, {Avruch}, {Batejat}, {B{\"a}hren}, {Bell}, {Bell}, {van
  Bemmel}, {Bennema}, {Bentum}, {Bernardi}, {Best}, {B{\^i}rzan}, {Bonafede},
  {Boonstra}, {Braun}, {Bregman}, {Breitling}, {van de Brink}, {Broderick},
  {Broekema}, {Brouw}, {Br{\"u}ggen}, {Butcher}, {van Cappellen}, {Ciardi},
  {Coenen}, {Conway}, {Coolen}, {Corstanje}, {Damstra}, {Davies}, {Deller},
  {Dettmar}, {van Diepen}, {Dijkstra}, {Donker}, {Doorduin}, {Dromer}, {Drost},
  {van Duin}, {Eisl{\"o}ffel}, {van Enst}, {Ferrari}, {Frieswijk}, {Gankema},
  {Garrett}, {de Gasperin}, {Gerbers}, {de Geus}, {Grie{\ss}meier}, {Grit},
  {Gruppen}, {Hamaker}, {Hassall}, {Hoeft}, {Holties}, {Horneffer}, {van der
  Horst}, {van Houwelingen}, {Huijgen}, {Iacobelli}, {Intema}, {Jackson},
  {Jelic}, {de Jong}, {Juette}, {Kant}, {Karastergiou}, {Koers}, {Kollen},
  {Kondratiev}, {Kooistra}, {Koopman}, {Koster}, {Kuniyoshi}, {Kramer},
  {Kuper}, {Lambropoulos}, {Law}, {van Leeuwen}, {Lemaitre}, {Loose}, {Maat},
  {Macario}, {Markoff}, {Masters}, {McFadden}, {McKay-Bukowski}, {Meijering},
  {Meulman}, {Mevius}, {Middelberg}, {Millenaar}, {Miller-Jones}, {Mohan},
  {Mol}, {Morawietz}, {Morganti}, {Mulcahy}, {Mulder}, {Munk}, {Nieuwenhuis},
  {van Nieuwpoort}, {Noordam}, {Norden}, {Noutsos}, {Offringa}, {Olofsson},
  {Omar}, {Orr{\'u}}, {Overeem}, {Paas}, {Pandey-Pommier}, {Pandey}, {Pizzo},
  {Polatidis}, {Rafferty}, {Rawlings}, {Reich}, {de Reijer}, {Reitsma},
  {Renting}, {Riemers}, {Rol}, {Romein}, {Roosjen}, {Ruiter}, {Scaife}, {van
  der Schaaf}, {Scheers}, {Schellart}, {Schoenmakers}, {Schoonderbeek},
  {Serylak}, {Shulevski}, {Sluman}, {Smirnov}, {Sobey}, {Spreeuw}, {Steinmetz},
  {Sterks}, {Stiepel}, {Stuurwold}, {Tagger}, {Tang}, {Tasse}, {Thomas},
  {Thoudam}, {Toribio}, {van der Tol}, {Usov}, {van Veelen}, {van der Veen},
  {ter Veen}, {Verbiest}, {Vermeulen}, {Vermaas}, {Vocks}, {Vogt}, {de Vos},
  {van der Wal}, {van Weeren}, {Weggemans}, {Weltevrede}, {White}, {Wijnholds},
  {Wilhelmsson}, {Wucknitz}, {Yatawatta}, {Zarka}, {Zensus}, \& {van
  Zwieten}}]{vanHaarlem2013A&A...556A...2V}
{van Haarlem} M.~P. {et~al.}, 2013, \aap, 556, A2

\bibitem[{{Visbal} {et~al}\mbox{.}(2015){Visbal}, {Haiman}, \&
  {Bryan}}]{Visbal2015MNRAS.453.4456V}
{Visbal} E., {Haiman} Z., {Bryan} G.~L., 2015, \mnras, 453, 4456

\bibitem[{Wilkins {et~al}\mbox{.}(2019)Wilkins, Lovell, \&
  Stanway}]{Wilkins2019arXiv191005220W}
Wilkins S.~M., Lovell C.~C., Stanway E.~R., 2019, \mnras, 490, 5359–5365

\bibitem[{Wise \& Abel(2007)}]{Wise2007}
Wise J.~H., Abel T., 2007, \apj, 671, 1559

\bibitem[{{Wise} {et~al}\mbox{.}(2012){Wise}, {Turk}, {Norman}, \&
  {Abel}}]{Wise2012ApJ...745...50W}
{Wise} J.~H., {Turk} M.~J., {Norman} M.~L., {Abel} T., 2012, \apj, 745, 50

\bibitem[{{Wolcott-Green} {et~al}\mbox{.}(2011){Wolcott-Green}, {Haiman}, \&
  {Bryan}}]{Wolcott-Green2011MNRAS.418..838W}
{Wolcott-Green} J., {Haiman} Z., {Bryan} G.~L., 2011, \mnras, 418, 838

\bibitem[{{Wouthuysen}(1952)}]{Wouthuysen1952AJ.....57R..31W}
{Wouthuysen} S.~A., 1952, \aj, 57, 31

\bibitem[{{Wyithe} \& {Loeb}(2013)}]{Wyithe2013MNRAS.428.2741W}
{Wyithe} J. S.~B., {Loeb} A., 2013, \mnras, 428, 2741

\bibitem[{{Xu} {et~al}\mbox{.}(2016){Xu}, {Wise}, {Norman}, {Ahn}, \&
  {O'Shea}}]{Xu2016ApJ...833...84X}
{Xu} H., {Wise} J.~H., {Norman} M.~L., {Ahn} K., {O'Shea} B.~W., 2016, \apj,
  833, 84

\bibitem[{{Yoshida} {et~al}\mbox{.}(2006){Yoshida}, Omukai, Hernquist, \&
  Abel}]{Yoshida2006}
{Yoshida} N., Omukai K., Hernquist L., Abel T., 2006, \apj, 652, 6

\bibitem[{{Yoshida} {et~al}\mbox{.}(2003){Yoshida}, {Sokasian}, {Hernquist}, \&
  {Springel}}]{Yoshida2003ApJ...598...73Y}
{Yoshida} N., {Sokasian} A., {Hernquist} L., {Springel} V., 2003, \apj, 598, 73

\bibitem[{{Yung} {et~al}\mbox{.}(2019){Yung}, {Somerville}, {Popping},
  {Finkelstein}, {Ferguson}, \& {Dav{\'e}}}]{Yung2019MNRAS.490.2855Y}
{Yung} L.~Y.~A., {Somerville} R.~S., {Popping} G., {Finkelstein} S.~L.,
  {Ferguson} H.~C., {Dav{\'e}} R., 2019, \mnras, 490, 2855

\bibitem[{{Zahn} {et~al}\mbox{.}(2011){Zahn}, {Mesinger}, {McQuinn}, {Trac},
  {Cen}, \& {Hernquist}}]{Zahn2011MNRAS.414..727Z}
{Zahn} O., {Mesinger} A., {McQuinn} M., {Trac} H., {Cen} R., {Hernquist} L.~E.,
  2011, \mnras, 414, 727

\bibitem[{{Zygelman}(2005)}]{Zygelman2005ApJ...622.1356Z}
{Zygelman} B., 2005, \apj, 622, 1356

\end{thebibliography}


\bsp
\label{lastpage}
\end{document}